\documentclass[12pt,preprint]{aastex}
\newcommand{\fxfopt}{$F_{\rm X}/F_{\rm opt}$}
\newcommand{\fxfi}{$F_{\rm X}/F_{\rm I}$}

\begin{document}

\title{``Hidden'' Seyfert 2 Galaxies in the Chandra Deep Field North}

\author{Carolin N.\ Cardamone\altaffilmark{1} and Edward C.\ Moran}
\affil{Astronomy Department, Wesleyan University, Middletown, CT 06459
       \\ \medskip{\rm and}\\}

\author{Laura E.\ Kay}
\affil{Department of Physics and Astronomy, Barnard College, 3009 Broadway,
       New York, NY 10027}

\altaffiltext{1}{Present address: Department of Astronomy, Yale University,
                 P.O.\ Box 208101, New Haven, CT 06520.}

\begin{abstract}
We have compared the X-ray--to--optical flux ratios (\fxfopt )
of absorbed active galactic nuclei (AGNs) in the {\it Chandra\/} Deep Field
North (CDF-N) with those of nearby, optically classified Seyfert~2 galaxies.
The comparison provides an opportunity to explore the extent to which the
local population of absorbed AGNs can account for the properties of the
distant, spectroscopically ambiguous sources that produce the hard X-ray
background.  Our nearby sample consists of 38 objects that well represent
the local Seyfert~2 luminosity function.  Integrated {\sl UBVRI\/} photometry
and broadband X-ray observations are presented.  Using these data, we have
simulated the \fxfopt\ ratios that local Seyfert~2s would exhibit if they
were observed in the redshift range $0.2 \le z \le 1.3$ as part of the CDF-N.
In the simulations we account for the effects of redshift on flux measurements
in fixed observed-frame bands, and the way the luminosity function of a given
population is sampled in a flux-limited survey like the CDF-N.  Overall, we
find excellent agreement between our simulations and the observed distribution
of \fxfopt\ ratios for absorbed AGNs in the CDF-N.  Our analysis has thus
failed to reveal any physical differences between the local population of
Seyfert~2s and CDF-N sources with similar X-ray properties.  These results
support the hypothesis that the nuclear emission lines of many distant hard
X-ray galaxies are hidden in ground-based spectra due to a combination of
observational effects: signal-to-noise ratio, wavelength coverage, and dilution
by host-galaxy light.

\end{abstract}


\section{Introduction}

Broadband X-ray observations have revealed that many active galactic nuclei
(AGNs) are heavily obscured by dense gas and dust located in their host
galaxies (e.g., Awaki et al.\ 1991).  The selective absorption caused by the
obscuring medium flattens (or inverts) the intrinsically steep X-ray spectra
of these AGNs, making them attractive candidates for the origin of the hard
(2--10 keV) X-ray background (XRB; Setti \& Woltjer 1989).  Detailed models
based on the observed properties of nearby AGNs have demonstrated that a
distant population of obscured objects is indeed capable of accounting for
the spectrum and intensity of the hard XRB (Comastri et al.\ 1995; Gilli,
Salvati, \& Hasinger 2001; Moran et al.\ 2001).  Consistent with this
expectation, the X-ray
colors of sources detected in extremely deep exposures with the {\it Chandra
X-ray Observatory}, which have resolved the majority of the hard XRB, indicate
that obscured AGNs are the most prevalent sources at faint hard X-ray fluxes
(Alexander et al.\ 2003).

Locally, the vast majority of obscured AGNs have the optical spectra of
Seyfert~2 galaxies, which are characterized by strong, narrow emission lines.
Spectroscopy of faint, hard {\it Chandra\/} sources should, therefore, provide
a straightforward means of confirming the Seyfert~2 model for the XRB.  But
a different picture has emerged: Follow-up studies of distant {\it Chandra\/}
sources have instead revealed a significant population of apparently
{\it normal\/} galaxies whose starlight-dominated optical spectra have only
weak emission lines, if any (e.g., Mushotzky et al.\ 2000; Barger et al.\
2001a, 2001b, 2002; Szokoly et al.\ 2004). Many such sources have the X-ray
properties of Seyfert~2 galaxies, but they seem to lack the corresponding
optical emission-line signatures.

There are several viable explanations for the normal optical appearance of
distant absorbed AGNs.  One possibility is that moderately luminous AGNs in
the past tend to be more heavily obscured than similar objects in the local
universe (Barger et al.\ 2001a, 2005).  A higher covering factor of the nuclear
obscuration would reduce the illumination of the narrow emission-line region
by the ionizing continuum, resulting in weaker narrow lines.  Alternatively,
extranuclear dust may play a greater role in obscuring our view of the
narrow emission-line regions of distant objects (Rigby et al.\ 2006).  Yet
another possibility is that distant AGNs may accrete predominantly in a
radiatively inefficient mode, whereby they produce significant hard X-ray
emission but far less of the soft X-ray and ultraviolet flux that is chiefly
responsible for the ionization of the narrow-line gas (Yuan \& Narayan 2004).

As an alternative to these scenarios,
Moran, Filippenko, \& Chornock (2002) have suggested that the limitations of
ground-based observing may be the culprit.  The small angular sizes
of distant sources cause their ground-based spectra to be dominated by light
from stars and/or \ion{H}{2} regions in the host galaxy, which can mask the
emission lines associated with their nuclear activity.  Integrated spectra of
local Seyfert~2s confirm that host-galaxy dilution would alter many of their
spectroscopic classifications if they were observed at modest redshifts with
ground-based facilities (Moran et al.\ 2002).  Still, the extent to which this
dilution affects the demographics of the distant X-ray galaxy population has
yet to be demonstrated.  Ultimately, a determination of whether the optically
normal appearance of distant absorbed AGNs is largely physical or observational
in origin has important implications for the nature of supermassive black
holes and their environments at earlier epochs.

Unfortunately, distant X-ray galaxies tend to be faint at all wavelengths,
which limits the amount and quality of information we have about their
properties.  For example, over half of the X-ray sources detected in the
2~Ms {\it Chandra\/} Deep Field North (CDF-N; Alexander et al.\ 2003; Barger
et al.\ 2003) have optical counterparts that are fainter than $R = 23$.
Clearly, high-quality optical spectra can only be obtained for the small
fraction of relatively bright sources in that field.  On the other hand,
broadband magnitudes and colors have been measured for nearly all of the
CDF-N sources.  X-ray--to--optical flux ratios (\fxfopt), therefore, offer
one of the best handles we have on the nature of these objects.  It has been
shown that the \fxfopt\ ratio broadly discriminates between various classes
of celestial X-ray sources (e.g., Stocke et al.\ 1991), in particular,
between luminous AGNs and truly normal (or quiescent) galaxies.  Thus, a
comparison of the \fxfopt\ ratios of the optically normal, X-ray--bright
objects that have turned up in the deep {\it Chandra\/} surveys to those
of local active galaxies with similar high-energy properties could be very
informative.  For instance, if host-galaxy dilution is generally not a
factor, we might expect the deficit of nuclear emission (line and continuum)
in the absorption or accretion-mode scenarios described above to lead to
systematically higher \fxfopt\ ratios in the distant population.

A fair comparison of the \fxfopt\ ratios of nearby and distant objects
requires the consideration of several important factors.  First, samples of
local and high-redshift AGNs are typically defined in very different ways.
Nearby samples contain objects recognized as AGNs for a variety of reasons
(e.g., X-ray brightness, strength of their emission in some other region of
the spectrum, optical emission-line properties, etc.) whereas distant X-ray
galaxies are usually identified on the basis of a sole property: detection
as an X-ray source.  In addition, the volume surveyed in flux-limited studies
such as the CDF-N is a sharp function of luminosity, which leads to a deficit
of low-luminosity sources and an over-representation of (rare) high-luminosity
objects in the derived source catalogs.  Thus, nearby and distant AGN samples
may contain inherently different types of objects and/or similar objects that
are drawn largely from different portions of the AGN luminosity function.
Another complication is that different portions of the rest-frame spectra
of nearby and distant galaxies fall within the fixed observed-frame bands
used to establish their \fxfopt\ ratios.  Redshift effects can have a
significant
impact on the perceived \fxfopt\ ratios of AGNs (Moran 2004; Peterson et al.\
2006) and must be accounted for.  And finally, only the integrated fluxes of
distant sources can be measured, and the same must be obtained for local
objects.

In this paper, we present a comparison of the \fxfopt\ ratios of absorbed AGNs
in the CDF-N with those of nearby galaxies classified optically as type~2
Seyferts.  Our approach accounts for the observational factors described
above by (1) employing a nearby sample that well represents the local
Seyfert~2 luminosity function and (2) accurately simulating how the nearby
objects would appear if they were observed in the CDF-N, including the effects
of how pencil-beam surveys like the CDF-N sample the luminosity function of a
given population.  This allows us to examine in detail the extent to which
nearby, well-characterized AGNs can explain the properties of distant,
spectroscopically ambiguous X-ray galaxies.  The criteria used to define
the comparison sample of absorbed AGNs from the CDF-N are outlined in \S~2.
In \S~3, the local Seyfert~2 sample is described, along with the integrated
optical and X-ray data we have collected for the objects.  Our simulations
are presented in \S~4, along with discussion of how the \fxfopt\ ratios of
Seyfert~2 galaxies are transformed by redshift and sampling effects.  Our
findings are summarized in the final section.

\section{The CDF-N Sample of Absorbed AGNs}

Our investigation of the \fxfopt\ ratios of absorbed AGNs requires an
appropriate sample of distant X-ray galaxies from a well-characterized survey,
and an unbiased sample of local objects with broadband X-ray and optical data.
For the distant X-ray galaxy sample, the 2~Ms CDF-N is an ideal resource. The
details of the {\it Chandra\/} observations and parameters of the $> 500$
sources detected in the survey have been thoroughly documented (Alexander et
al.\ 2003).  In addition, deep optical imaging and spectroscopy of the sources
have been obtained with the Subaru 8~m and Keck 10~m telescopes (Barger et al.\
2002, 2003), yielding optical fluxes and, for many objects, spectroscopic
redshifts.

The sources we have selected from the CDF-N have X-ray properties similar to
those of nearby Seyfert~2s and are drawn from a well-defined portion of the
deep survey.  First, we select only sources with total exposure times between
1.5~Ms and 2.0~Ms.  This exposure time range brackets the strong peak in the
CDF-N source exposure time distribution centered at 1.7~Ms (Alexander et al.\
2003), and because it is narrow, it allows us to establish an effective X-ray
flux limit and solid angle for the deep survey, which are required for the
simulations described below.  Next, since we are chiefly concerned with the
origin of the XRB, we select CDF-N sources with 2--8~keV hard-band detections
and absorbed X-ray spectra with effective photon indices $\Gamma<1.5$ (as
indicated by their ``hardness ratios''). These are the sources responsible for
the hard XRB, and based on observations of nearby objects, they are expected
to be Seyfert~2 galaxies.  Finally, we require that the included sources have
a measured spectroscopic redshift.

Over 80\% of the sources that satisfy these criteria have redshifts between
$z = 0.2$ and $z = 1.3$.  We have further restricted our CDF-N sample to this
redshift range for two reasons.  First, objects closer than $z \approx 0.2$
are probably extended in optical images, and published magnitudes for them
may not reflect their total optical emission.  Second, our simulations (\S~4)
employ {\sl UBVRI\/} data for nearby Seyfert~2s to yield the observed-frame
$I$-band fluxes they would have at various redshifts.  At $z = 1.3$, the
rest-frame $U$ band is roughly centered on the observed $I$ band.  Adopting
this redshift limit thus eliminates the need for significant extrapolation
of our local galaxy spectra to wavelengths shortward of $U$.

A total of 59 CDF-N sources meet all of our selection criteria.  Using
published 2--8~keV fluxes and $I$-band magnitudes (Alexander et al.\ 2003;
Barger et al.\ 2002, 2003), we have computed their observed-frame \fxfi\
flux ratios.  Optical spectra are published for only 38 of the objects
(Barger et al.\ 2002), but a visual inspection of these indicates only half
a dozen or so clearly have the spectral signatures of narrow-line AGNs.
Curiously, one other object is reported to have broad emission lines, though
they appear to be weak in the Barger et al.\ data.  The spectra of the rest
of the objects are consistent with those of normal galaxies, or are ambiguous
because of the signal-to-noise ratio and/or wavelength coverage of the data.
Presuming the rest of the objects we have selected to be similar, it is safe
to conclude that the majority of the absorbed AGNs in our CDF-N sample are
not considered to be Seyfert~2s on the basis of their ground-based optical
spectra.  The redshifts, {\it Chandra\/} exposure times, 2--8 keV fluxes,
and 2--8 keV luminosities of the CDF-N objects are shown in Figure~1.

\section{The Local Sample of Seyfert 2 Galaxies}

To ensure that our comparison of the \fxfopt\ ratios of nearby and distant
objects is fair, it is vital that we employ a local sample 
that is as complete and unbiased as possible.  However, because of the
variety of ways in which Seyfert~2s have been discovered and the fact that
their luminosity function is not firmly established, this is a
non-trivial matter.  The biases that result when samples are flux-limited
and defined on the basis of a particular property (e.g., ultraviolet excess
or far-infrared color) are well documented (Ho \& Ulvestad 2001).  In
addition, samples of Seyfert~2 galaxies can be tainted by spectroscopic
misclassifications.

To minimize the effects of selection biases and contamination in our study,
we have chosen to use objects drawn from the distance-limited sample of
Ulvestad \& Wilson (1989; hereafter UW89), which consists of all Seyfert
galaxies known (at the time of its definition) with redshifts $cz \leq 4600$
km~s$^{-1}$ ($z \leq 0.0153$) and a declinations $\delta \geq -45^{\circ}$.  
Because the objects were included on the basis of distance, and not some
observed property, and because their nuclear activity was noticed for a
variety of reasons, the sample is free of significant selection biases.
In addition, detailed optical investigations of this sample have verified
that all 31 of the Seyfert~2s it contains are bona fide narrow-line AGNs
(Moran et al.\ 2000).  For this study, we also include the 7 objects listed
by UW89 as ``narrow-line X-ray galaxies'' (NLXGs), despite the fact that
several of them are technically intermediate type~1 Seyferts that display
weak, broad H$\alpha$ components in high-quality optical spectra.  Our
analysis of {\sl ASCA\/} data for the NLXGs (\S~3.2) has confirmed that
all of the objects are absorbed X-ray sources, with column densities of
$\sim 10^{22}$ cm$^{-2}$ or more.  Thus, over a range of redshifts they
would satisfy the spectral flatness criterion used above to select absorbed
AGNs in the CDF-N (\S~2).  Including the NLXGs, our local sample of absorbed
AGNs (which we refer to as ``Seyfert~2s'' for convenience) stands at 38
objects.

We note that not every galaxy within the UW89 distance and declination limits
has been searched for a Seyfert nucleus, so the sample must be incomplete
to some degree.  Indeed, some Seyfert galaxies have been discovered within
the sample volume since 1989.  The level of incompleteness is probably most
significant at low values of the nuclear luminosity, where, in many cases,
an accurate emission-line classification cannot be made without careful
starlight template subtraction (Ho, Filippenko, \& Sargent 1997).  Still,
several lines of
evidence suggest that the UW89 sample, while falling short of perfection,
is nonetheless a very good one.  First, the radio properties of the UW89
Seyferts are broadly consistent with those of objects in other samples, e.g.,
the CfA sample (Kukula et al.\ 1995).  Second, as Figure~3 of Lumsden \&
Alexander (2001) illustrates, the UW89 sample extends to much lower
luminosities than other well-studied collections of Seyfert~2s, such as
the CfA/12~$\mu$m (Tran 2001) and {\sl IRAS\/} (Lumsden et al.\ 2001) samples.
Thus, it contains more typical Seyfert~2s and suffers less from an
over-representation of high-luminosity objects than these other samples.
Finally, the X-ray luminosity density of the Seyfert~2 population derived
from the UW89 sample is able to account for
both the intensity and spectral slope of the 2--10 keV X-ray background
(Moran et al.\ 2001).  Taken in combination, these results suggest that the
UW89 sample must represent the luminosity function of type~2 Seyfert galaxies
reasonably well.

\subsection{Broadband Optical Data}

The fluxes measured for distant CDF-N objects reflect their integrated optical
and X-ray emission; comparable data are needed for local Seyfert~2 galaxies so
that we can simulate what their \fxfopt\ ratios would be if they were observed
at modest redshift in the CDF-N.  Surprisingly, although the UW89 objects are
among the most well-studied Seyfert~2 galaxies, relatively little information
about their integrated optical fluxes has been published. Integrated magnitudes
in the blue and visual bands can be found for about 60\% of the sample, and
data at redder wavelengths are even more sparse.  In this section we describe
our {\sl UBVRI\/} observations of over half of the UW89 sample, and our methods
of determining integrated magnitudes for the remainder of the objects.

\subsubsection{UBVRI Observations}

Our optical data were acquired with the 0.9-m WIYN telescope at Kitt Peak and
the 1.3-m McGraw Hill telescope at the MDM Observatory during six separate
runs between 2003 January and 2006 January.  On the WIYN telescope, we used
the $2048 \times 2048$ S2KB CCD, which affords a $\sim$ $20' \times 20'$ field
of view and an image scale of $0\farcs6$ per pixel.  At MDM we employed the
$1024 \times 1024$ ``Templeton'' CCD, which has an $8\farcm5 \times 8\farcm 5$
field of view and a scale of $0\farcs5$ per pixel.  Images were obtained
with Harris {\sl UBVRI\/} filters on the 0.9-m, and on the 1.3-m, Harris
{\sl BVR\/} filters were used in conjunction with a Bessell $U$ filter and
an Arizona $I$ filter.  During our 2003 October, 2004 March, and 2006 January
runs, we obtained photometric observations in all five bands for 21 UW89
galaxies.

We processed our images using standard IRAF procedures. In each, an integrated
instrumental magnitude was measured within a circular aperture centered on the
nucleus of the UW89 galaxy.  The size of the aperture was initially selected
to include all of the galaxy flux visible when the image was displayed with
extreme settings of the contrast and dynamic range.  The aperture was then
resized to the point at which increasing its radius did not result in an
increase of the galaxy's flux.  The flux from foreground stars was
measured and subtracted from the total flux in the galaxy aperture.  The sky
background level was estimated within a concentric annulus placed well outside
the galaxy aperture.  A few of the UW89 objects have nearby companions, which
have not been excluded in our measurements.  Our objective is to compare the
UW89 sample and distant X-ray galaxies, and the flux from companions would
not be separable in observations of the latter.  By including companion flux
in the nearby sample we preserve the true diversity of the morphologies and
integrated colors of its members, making our comparison as valid as possible.
In the end, this concerns only a handful of objects: NGC~5929 is interacting
with NGC~5930, a spiral galaxy of comparable brightness; NGC~262 has a minor
companion, LEDA 212600, and two fainter satellite galaxies; and NGC~1667 has
a single dwarf companion.

We observed equatorial standard star fields (Landolt 1992) to calibrate our
instrumental magnitudes. Average extinction coefficients for each band were
obtained from Landolt (1983).  Following Bessell (1995), we transformed the
magnitudes of the standard stars to the Johnson-Kron-Cousins system defined
by Bessell (1990). The formal uncertainties in our {\sl UBVRI\/} measurements,
which include the effects of photon statistics, flat-fielding accuracy,
aperture size, and transformation to the standard scale, are estimated to be
0.02--0.04 mag.  Table~1 lists the {\sl UBVRI\/} photometry results for the
21 objects we observed.

For 12 of these galaxies, integrated photoelectric photometry in $B$ and
$V$ (with typical uncertainties of 0.1--0.2 mag) is published in the Third
Reference Catalog of Bright Galaxies (RC3; de Vaucouleurs et al.\ 1991).
This provides a limited opportunity to check the accuracy of our measurements.
As the left panel of Figure~2 indicates, the differences between our $V$-band
magnitudes
and those listed in the RC3 are (for 11 objects) consistent with the expected
uncertainties in the two quantities (the median offset is 0.04~mag with a
standard deviation of 0.10~mag).  One significant discrepancy was uncovered,
however:\ We find Mrk~3 to be brighter than the RC3 values by 0.58~mag in $V$
and 0.43~mag in $B$.  The problem may be related to a very bright foreground
star located within the aperture we used to extract the galaxy's flux.  The
star is {\it not\/} responsible for our flux being too high --- using an
aperture that completely excludes the star we obtain a $B$ value that
is fainter by only 0.2~mag.  Thus, we are confident that we have successfully
removed the contribution of the star in our full-aperture data, but it is not
clear how the contamination was handled in the RC3 measurement.  We note that
the RC3 also lists values of $m_B$ --- photographic magnitudes from the
Shapley-Ames catalog (Sandage \& Tammann 1981) that have been reduced to the
$B_T$ system.  The $m_B$ value for Mrk~3 of $13.55\pm 0.17$ is
entirely consistent with our measurement of $B = 13.60$.  Thus, given the
overall agreement between our measurements and those listed in the RC3, we
have confidence in the accuracy of our photometry.

\subsubsection{Integrated Magnitudes of the Unobserved Galaxies}

Information about the integrated magnitudes of the 17 (mostly southern)
galaxies we did not observe is also available from the RC3 ($U_T$, $B_T$,
and $V_T$) and the ESO-Uppsala survey ($B_T$ and $R_T$; Lauberts \& Valentijn
1989).  We have adopted photoelectric magnitudes from the RC3 whenever they
are available (12 galaxies), and have supplemented these with photographic
$R_T$ magnitudes from the ESO catalog when $B_T$(ESO) agrees with $B_T$(RC3).
Three other objects that have only $m_B$ photographic magnitudes in the RC3
have $B$ and $R$ measurements in the ESO catalog; the $B$ magnitudes agree
in each case, so we have adopted the ESO values for these galaxies.  Only
$m_B$ data are available for the two remaining objects.

To estimate integrated magnitudes in the bands lacking published data, we have
used the $B - V$ and/or $B - R$ colors of the objects to determine the most
appropriate Johnson-Cousins color template from the compilation of Fukugita
et al.\ (1995).  The majority of the galaxies are best represented by an Sab
template, though for several an S0 (6 galaxies) or Sbc (2 galaxies) template
provides the closest match.  For the two objects with $m_B$ magnitudes only
(thus no integrated colors), we have adopted the Sab template.  In all cases,
the template we have selected is consistent with the galaxy's morphological
type listed in the NASA/IPAC Extragalactic Database (NED).

Turning once more to the 11 objects we observed that have reliable 
photoelectric data in the RC3, we have compared the $R$-band magnitudes
extrapolated from their $V_T$ values to the $R$ magnitudes that we derived
from our images.  As the right panel of Figure~2 indicates, the median
difference in these magnitudes is 0.04~mag with a standard deviation of
0.11~mag.  The similarity between the left and right panels of Figure~2
suggests that the application of a Fukugita et al.\ (1995) template
does not add an appreciable systematic error to the extrapolated magnitudes,
giving us confidence that the extrapolated magnitudes for the unobserved
objects are reasonably accurate.  The integrated magnitudes obtained from the
literature, together with those derived from application of the appropriate
color template, are listed in Table~2.  The final column of Table~2 indicates
the bands for which published data are available and the Fukugita et al.\
(1995) galaxy template that was used.

\subsubsection{Corrections for Galactic Extinction}

Corrections for Galactic extinction are necessary for a determination of the
true optical fluxes of the UW89 Seyfert~2s and for comparisons to galaxies in
other samples.  We corrected our magnitudes by computing $A_{\lambda}$ for
each object in each of the five bands.  Values of the color excess $E(B-V)$
in the direction of each galaxy (originating from Schlegel et al.\ 1998)
were obtained from NED.  The total absorption in each band was then calculated
from $A_{\lambda}/E(B-V)$ using Table~6 of Schlegel et al.\ (1998).  The final
extinction-corrected magnitudes for all 38 UW89 objects are listed in Table~3.
For clarity, magnitudes obtained from our observations or the RC3 are given
in plain type and those derived with the use of a Fukugita template
are given in italics.  The Galactic latitudes of the UW89 members span a wide
range, so the extinction corrections vary considerably from object to object.

\subsubsection{Ultraviolet Data}

In the simulations described in \S~4, information about the slope of
the near-UV spectra of the UW89 objects is needed to ensure that the
observed-frame $I$-band fluxes we predict for them are accurate for all
assumed redshifts up to our limit of $z = 1.3$.  The atlas of galaxies
observed with {\sl GALEX\/} (Gil de Paz et al.\ 2006) provides integrated
fluxes for 8 UW89 Seyfert~2s (Mrk~3, NGC 262, 1068, 1386, 2992, 4117,
4388, and 7582) at a near-UV wavelength of 2267~\AA .  After correcting
for Galactic extinction [$A_{\rm NUV} = 8 \times E(B-V)$; Gil de Paz et
al.\ 2006], we find that the ${\rm NUV} - U$ colors of our 8 objects range
from 1.64 to 3.17.  For the rest of the UW89 objects, we adopt the median
value of ${\rm NUV} - U = 2.12$.  

\subsection{X-ray Data}

Broadband X-ray data in the $\sim 0.5$--10 keV energy range are available
for the entire UW89 Seyfert~2 sample.  Nearly all (36/38) of the objects
were observed with the {\sl ASCA\/} satellite; the two remaining galaxies
(NGC~5283 and NGC~5728) have been observed with the {\it Chandra X-ray
Observatory}.  Although several other UW89 galaxies have also been observed
with {\it Chandra}, the {\sl ASCA\/} data are preferred because of the
consistent depth of the exposures and the fact that, due to the large
{\sl ASCA\/} beam ($\sim 3$ arcmin half-power diameter), we can be sure
that they represent the total X-ray flux from the nucleus and host galaxy.
The nucleus is likely to dominate in the majority of cases, at least at
the higher X-ray energies, but many objects are weak and their extended
X-ray emission (from supernova remnants, X-ray binaries, etc.) could be
comparable to the nuclear flux.

Details of the {\sl ASCA\/} observations and data reduction for the UW89
sample have been described by Moran et al.\ (2001); a brief summary is provided
here. The data were obtained from our own observations and from the HEASARC
data archive at NASA's Goddard Space Flight Center.  The {\sl ASCA\/} exposure
times of the UW89 Seyfert~2s are uniformly long (most are in the 35--45 ks
range), and the targets were placed at the ``1-CCD'' off-axis position in most
of the images.  For this work we focus on data collected with the Gas Imaging
Spectrometers (GIS) on board {\sl ASCA}; compared to the satellite's SIS
instruments, the GIS have better hard X-ray sensitivity and more consistent
response, and due to their larger field of view, background estimation is
more straightforward with them.

The {\it Chandra\/} images of NGC~5283 and NGC~5728 were obtained from the
data archive at the Chandra X-ray Observatory Center (CXC).  The objects were
observed with the ACIS-S instrument for 9.8 ks and 19.0 ks, respectively.
Both sources are relatively weak ($\sim 0.06$ count~s$^{-1}$), so spectral
distortions resulting from photon pile-up are not a concern.

We extracted source and background events for all of the {\sl ASCA\/} and
{\it Chandra\/} data sets, and generated response and effective area files
specific to the individual observations.  All 38 UW89 Seyfert~2s were detected
above a signal-to-noise ratio of 4 (full band).  For 25 objects, the net counts
obtained were sufficient to allow spectral modeling with the XSPEC software
(Arnaud 1996).  We have modeled the spectra as the sum of three components: a
weakly absorbed power law with a photon index $\Gamma_1$ and associated column
density of $(N_{\rm H})_1$, a heavily absorbed power law with slope $\Gamma_2$
and column density $(N_{\rm H})_2$, and a Gaussian Fe~K$\alpha$ line of width
$\sigma_{K\alpha}$ centered at energy $E_{K\alpha}$.  In all instances but one,
an acceptable fit with reasonable best-fit spectral parameters was obtained.
The exception is NGC~1068, which has a far more complex broadband X-ray
spectrum (Iwasawa, Fabian, \& Matt 1997; Matt et al.\ 1997).  Table~4 lists the
adopted distances to the galaxies (see Moran et al.\ 2001), the X-ray spectral
parameters derived from our fits, and the associated X-ray fluxes in the
0.5--2 keV and 2--8 keV bands.  We note that while our relatively simple
spectral models generally afford statistically acceptable fits, they may not
represent the best physical description of the X-ray emission in every case.
The main purpose of our spectral analysis is to provide accurate fluxes,
which it does.  This is true even for NGC~1068; our approach yields soft-
and hard-band fluxes that are respectively within 10\% and 1\% of those
obtained using a more complex model that provides a good fit.

For the 13 weakly detected objects, X-ray fluxes were estimated from ratios
of the counts detected in hard (4--10 keV) and soft (1--4 keV) bands.  First,
we computed the median Seyfert~2 X-ray spectrum based on the spectral fits
obtained for the 25 ``strong'' sources above.  The median spectrum is
characterized by the following parameters: $\Gamma_1 = 1.78$, $\Gamma_2 =
1.70$, $(N_{\rm H})_1 = 0$, and $(N_{\rm H})_2 = 2.42 \times 10^{23}$
cm$^{-2}$.  (An Fe~K$\alpha$ component is not included, for reasons that
will become clear below.)  The median model is similar to the composite
Seyfert~2 X-ray spectrum derived from the summed emission of the UW89
objects (Moran et al.\ 2001), despite the fact that the latter is
dominated by the most luminous sources.

To estimate the X-ray fluxes of the weak sources, we fixed the parameters
of the median model and varied the normalizations of the two power-law
components in XSPEC until the hard-to-soft counts ratio associated with the
model matched the observed counts ratio.  We then fixed the ratio of the
normalizations and scaled them until the count rate implied by the model
was identical to the total observed count rate.  The fluxes in 0.5--2.0 keV
and 2.0--8.0 keV ranges were then computed from the model.  To validate
our approach, we applied the same procedure to the ``strong'' sources whose
spectra could be modeled directly.  As Figure~3 indicates, the 2--8 keV
fluxes obtained directly from spectral fitting and those obtained using the
median model differ by only a few percent in most cases (even {\it without\/}
a contribution from an Fe~K$\alpha$ line).  Based on this good agreement,
we are confident that the fluxes we have derived for the 13 weak objects
(also listed in Table~4) are reasonably accurate.

\section{Simulations}

\subsection{Approach}

A direct, fair comparison of the \fxfopt\ ratios of distant absorbed AGNs
with those of nearby Seyfert~2 galaxies cannot be made.  First of all, the
\fxfopt\ ratio is measured in the observed frame, so its value for a given
object varies with redshift.  Secondly, the ways in which samples of distant
and nearby sources are assembled naturally lead to different luminosity
distributions in the samples, which in turn affect the distributions of their
\fxfopt\ ratios.  Our approach, therefore, is to take a sample of nearby
Seyfert~2s that well represents the local luminosity function and simulate the
distribution of flux ratios that would result if they were observed under
the same conditions (and with the same redshift distribution) as the distant
sources.  This minimizes the effects of redshift and selection bias.

We begin by applying the information listed in Table~4 to determine the fluxes
of each UW89 source in the observed 0.5--2 keV and 2--8 keV bands as a function
of redshift.  The luminosity distances used in the calculations are based on an
$H_0$ = 70 km~s$^{-1}$~Mpc$^{-1}$, $\Omega_{\rm M} = 1/3$,  $\Omega_{\Lambda}
= 2/3$ cosmology.  The results establish the redshift range within which each
UW89 object would be detectable if observed as part of the $t \ge 1.5$~Ms
portion of the CDF-N survey.  Specifically, we apply the same criteria used to
define our CDF-N sample of absorbed AGNs: a 2--8 keV flux limit corresponding
to the this exposure time ($1.8 \times 10^{-16}$ erg~cm$^{-2}$~s$^{-1}$) and
an effective spectral index $\Gamma \le 1.5$, which corresponds to a flux ratio
$F_{\rm 2-8}/F_{\rm 0.5-2} \ge 2$.  It is interesting to note that, based on
these criteria, four UW89
galaxies would not be included in the CDF-N (as absorbed AGNs) in the $0.2 \le
z \le 1.3$ range.  The spectra of NGC~1068, NGC~1386, and NGC~5135 are too
steep to meet the spectral flatness criterion while their 2--8 keV fluxes are
above the hard X-ray flux limit.  The fourth object, NGC~4941, falls below the
flux limit before $z = 0.2$.  Only 10 objects would be detectable in the CDF-N
all the way out to our redshift limit of $z = 1.3$.

We use Monte Carlo methods to simulate the \fxfi\ distribution that nearby
Seyfert~2 galaxies would have if observed in the CDF-N, randomly selecting
a redshift (weighted by the CDF-N redshift distribution) and a UW89 galaxy 
(unweighted, since to first order the UW89 sample is the local Seyfert~2
luminosity function).  We first verify that the UW89 object would be included
in the CDF-N as an absorbed AGN at the chosen redshift.  If not, another
galaxy is selected at the same redshift.  Next, we determine the likelihood
that an object with the UW89 galaxy's X-ray luminosity would be included in
the CDF-N.  For this test, we have combined the CDF-N flux limit and survey
solid angle (170 arcmin$^2$ for $t = 1.5$~Ms) to estimate the volume searched
in the CDF-N as a function of minimum detectable 2--8~keV luminosity.  The
results are plotted in Figure~4, along with the fixed volume represented by
the UW89 sample (calculated by Moran et al.\ 2001).  Below a luminosity of
$\sim 3 \times 10^{42}$ erg~s$^{-1}$, the volume searched in the CDF-N is
less than that of the UW89 sample.  Therefore, in this $L_{\rm X}$ range, the
ratio of the CDF-N volume to the UW89 volume defines the probability that a
local object of a given luminosity would be included in the CDF-N.

If a UW89 galaxy passes all the above tests, the {\sl UBVRI\/} photometry
reported in \S~3 is used to compute its integrated, observed-frame $I$-band
flux.  The optical spectrum of the object is approximated by converting the
broadband magnitudes to flux densities at the band centers and assuming they
are joined by power laws.  The spectrum is shifted and dimmed appropriately
for the selected redshift; the portion falling within the observed $I$ band
is then integrated over the width of the band to give us the optical flux.
As the redshift approaches $z = 1.3$, the rest-frame UV spectrum shortward of
the center of the $U$ band enters the observed-frame $I$ band.  The NUV data
(\S~3.1.4) is used to extrapolate to shorter wavelengths, though the value
of the ${\rm NUV} - U$ color we adopt affects the flux by $< 1$\%.  Using the
derived optical flux, the \fxfi\ ratio of the object is then calculated.
The process continues until an \fxfi\ distribution composed of $10^4$ UW89
objects is obtained.

\subsection{Redshift and Sampling Effects}

Before presenting the results of our simulations and a comparison to the
CDF-N, we explore the way source redshifts and the flux-limited nature of
deep surveys combine to influence the \fxfopt\ ratios of a population of
absorbed AGNs.

As Table~3 indicates, the intrinsic integrated optical colors of the UW89
objects are quite red ($B - I \approx 2$).  In the X-ray band, the heavy
absorption in Seyfert~2 galaxies usually hardens their observed X-ray spectra
considerably (see the composite UW89 X-ray spectrum in Fig.~1 of Moran et
al.\ 2001).  Thus, as the redshift of a typical Seyfert~2 galaxy increases,
a brighter portion of its rest-frame X-ray spectrum is shifted into the
observed 2--8 keV band, and a fainter portion of its rest-frame optical
spectrum is shifted into the observed $I$ band.  The observed-frame \fxfopt\
ratio should therefore increase significantly with redshift.  This effect
is clearly demonstrated in Figure~5, where we have plotted $F_{\rm X}$,
$F_{\rm I}$, and \fxfi\ vs.\ redshift for four UW89 Seyfert~2s spanning a
wide range of intrinsic \fxfi\ ratios.  Between $z = 0$ and $z = 1.5$, the
observed flux ratios of these objects increase by factors of 15 to 35.

Of equal importance are the effects of sampling in a flux-limited survey
such as the CDF-N.  As Figure~4 illustrates, the volume searched for X-ray
galaxies in the CDF-N is a strong function of the observed 2--8 keV
luminosity.  This naturally leads to Malmquist bias in the CDF-N source
catalog, i.e., an under-representation of relatively abundant sources with
low X-ray luminosities, and an over-representation of rare, high-luminosity
sources.  If \fxfi\ happens to depend on $L_{\rm X}$ (and it does; see \S~4.3),
these Malmquist effects will be imprinted on the \fxfi\ distribution for
absorbed AGNs in the CDF-N.

In combination, the effects of redshift and sampling can dramatically alter
the observed \fxfi\ distribution for Seyfert~2 galaxies.  In Figure~6 we have
plotted the rest-frame \fxfi\ distribution for the UW89 sample, along with
the distribution obtained by simulating CDF-N observations of the UW89 objects
(as described in the previous section).  Clearly, the two distributions bear
no resemblance to each other, even though they are derived from the same set
of objects.  This illustrates why a direct comparison of the \fxfopt\ ratios
of nearby and distant sources would yield misleading results.  More generally,
Figure~6 indicates that \fxfopt , as an activity diagnostic, can be ambiguous.
X-ray survey results are often summarized with plots that compare the X-ray
and optical fluxes of the detected sources, with diagonal lines drawn for
constant values of \fxfopt\ (e.g., Alexander et al.\ 2003; Bauer et al.\ 2004).
Frequently, the region on these plots represented by
log \fxfopt\ $> -1$ are labeled ``AGNs,'' while that represented by log
\fxfopt\ $< -2$ are labeled ``galaxies.''  Our investigation of the UW89
sample reveals that Seyfert~2s at modest redshift can have \fxfopt\
ratios well outside the range expected for AGNs.

\subsection{Comparison to the CDF-N}

The \fxfi\ distribution for the absorbed AGNs in the CDF-N and the results
of our simulations are compared in Figure~7.  As the Figure indicates, the
two \fxfi\ distributions are broadly consistent with each other: they peak
at the same place and have roughly the same width.  The match is especially
good for values of log~\fxfi\ $\ge -1$. Note that the CDF-N distribution
comprises just 59 objects, so there is some statistical uncertainty associated
with the number of objects in each bin of that distribution.  The only
possible discrepancy occurs at the lowest \fxfi\ ratios, where the simulated
UW89 distribution falls consistently below the CDF-N distribution.  Given
the nature of the rest-frame \fxfi\ distribution of the UW89 sample (Fig.~6),
there is no chance that the good agreement between the CDF-N and simulated
UW89 flux-ratio distributions is coincidental.  Instead, it must be a
reflection of the similarity between the nearby and distant populations of
absorbed AGNs.

A more detailed comparison is provided in Figure~8, which plots the observed
\fxfi\ ratio as a function of observed 2--8 keV luminosity for the 59 CDF-N
sources and a UW89 simulation consisting of 75 successful trials.  Two things
are immediately obvious in Figure~8: (1) the CDF-N and UW89 points occupy
similar locations in the \fxfi\ -- $L_{\rm X}$ plane, and (2) \fxfi\ scales
roughly linearly with $L_{\rm X}$ for both populations, albeit with a fair
amount of dispersion. The fact that UW89 and CDF-N galaxies of a certain
nuclear luminosity ($L_{\rm X}$) have about the same range of \fxfi\ ratios
indicates that they are fundamentally similar objects in terms of their
optical properties.  Moreover, because the slope of the ``correlation''
between \fxfi\ and $L_{\rm X}$ is about unity, the median optical luminosity
of the objects must be roughly constant and independent of the luminosity of
the nucleus. In a statistical sense, therefore, it appears that the optical
luminosities of absorbed AGNs (with observed $L_{\rm X} \le 10^{43}$
ergs~s$^{-1}$) are dominated by emission from the host galaxy.

It is also evident from Figure~8 that the match between the UW89 and CDF-N
samples is not perfect.  In particular, there are no UW89 objects with
observed hard X-ray luminosities above $\sim 2 \times 10^{43}$ ergs~s$^{-1}$,
whereas the CDF-N sample extends to $L_{\rm X} = 10^{44}$ ergs~s$^{-1}$.
Also, there appear to be too few UW89 objects with $L_{\rm X} < 10^{42}$
ergs~s$^{-1}$ and log~\fxfi\ $< -1$, consistent with the discrepancy between
the flux-ratio distributions shown in Figure~7.  Although several factors
can affect the exact location of an absorbed AGN in the \fxfi\ -- $L_{\rm X}$
plane, the differences indicated in Figure~8 are almost certainly related to
the completeness of the UW89 sample rather than physical differences between
the nearby and distant sources.  At the high-$L_{\rm X}$ end, the volume
associated with the UW89 sample is too small to include rare, high-luminosity
objects, which are in fact over-represented in the CDF-N due to the large
volume it surveys for such sources (see Fig.~4).  At the low-$L_{\rm X}$
end, the incompleteness of the UW89 sample noted in \S~3 is probably the
primary issue.  Again, the nearby Seyfert~2 galaxies that are absent in
the UW89 sample are likely to be objects with low-luminosity nuclei, such
as those for which starlight template subtraction is required for an accurate
emission-line classification.  Having more local galaxies with low X-ray
luminosities would increase the representation of low-\fxfi\ sources in our
simulations.  The fact that our simulations reproduce the flux-ratio
distribution of distant absorbed AGNs as well as they do implies that the
shortcomings of the UW89 sample are not too severe.

\subsection{Optically Normal X-ray Galaxies in the Deep Surveys}

If nearby Seyfert~2 galaxies are able to account for the relative X-ray and
optical properties of distant absorbed AGNs, why do the objects detected in
the {\it Chandra\/} deep surveys often lack emission-line evidence for
nuclear activity?  As Moran et al.\ (2002) have demonstrated, a combination
of observational factors --- host-galaxy dilution, signal-to-noise ratio, and
wavelength coverage --- are capable of making many UW89 Seyfert~2s appear
``normal'' in their integrated spectra.  The bulk of the UW89 and CDF-N
objects overlap in terms of their optical luminosities, which in both cases are
dominated by host-galaxy emission, so these observational factors should affect
the ground-based spectra of distant absorbed AGNs in a similar way.  It would
seem, therefore, that additional hypotheses for the optically normal appearance
of the CDF-N population --- at least those objects that satisfy our selection
criteria --- are unnecessary at this time.

However, in a recent study of objects from the {\it Chandra\/} Deep Field South
(CDF-S), Rigby et al.\ (2006) have argued that the absence of strong AGN-like
emission lines in the ground-based spectra of distant X-ray galaxies results
primarily from obscuration of the narrow-line region by extranuclear dust,
rather than host-galaxy dilution.  Their conclusions are based on the finding
that the morphologies of optically active galaxies (with broad emission lines
or high-excitation narrow lines) and optically dull galaxies (with weak and/or
low-excitation emission lines) differ statistically.  Optically active galaxies
in the CDF-S tend to have high ratios of their semiminor and
semimajor axes $b/a$, whereas optically dull objects have a relatively flat
distribution of $b/a$.  Taking the measured axis ratio as a proxy for
inclination, Rigby et al.\ suggest that the optically dull sources are missing
AGN-like emission lines because extranuclear dust obscures the narrow-line
region in the more inclined galaxies.

To examine the inclination hypothesis, we have compared axis ratio
distributions for appropriate subsets of the Rigby et al.\ and UW89 samples.
For the CDF-S galaxies, we have compiled the $b/a$ distribution for those
objects in the $0.5 \le z \le 0.8$ subsample of Rigby et al.\ that satisfy
our X-ray selection criteria, i.e., detection in the 2--8~keV band and a
(2--8 keV)/(0.5--2 keV) flux ratio in excess of 2.  Although we have ignored
their emission-line classifications, this resticted CDF-S sample of 15 objects
includes just one optically active source --- a narrow-line object --- so any
possible confusion introduced by the presence of broad-line AGNs has been
eliminated.  Likewise, we have limited the UW89 comparison sample to include
only the 18 objects that would be detected in the CDF-S ($F_{\rm 2-8} > 3
\times 10^{-16}$ ergs~cm$^{-2}$~s$^{-1}$) at $z \ge 0.5$.  We have estimated
axis ratios for the UW89 galaxies using our images or images available from
NED. Our measurements, obtained both by hand and with the {\it ellipse\/} task
in IRAF, are based on the shape of the outer, low surface-brightness isophotes.
The two methods yield very similar results for all objects where both could
be successfully employed.  In a handful of cases (e.g., interacting galaxies)
the output from {\it ellipse\/} is suspect and we favor the values measured
by hand.  Our best estimates of $b/a$ for the UW89 subsample are listed in
Table~5.  As Figure~9 indicates, local Seyfert~2s have a very broad
distribution of $b/a$, implying that they are at least as inclined as the
optically dull objects in the CDF-S.  One caveat here is that $b/a$ has not
been measured in exactly the same way for the nearby and distant sources. 
However, given the coarse binning used in Figure~9 it is unlikely that a
different measurement approach for the local sample would alter these results
significantly.  The fact that the UW89 objects have strong nuclear emission
lines suggests that inclination, and the associated effects of extranuclear
dust, cannot be the primary origin of the optically normal appearance of the
distant, absorbed X-ray galaxies.

In a recent complementary study, Peterson et al.\ (2006) have examined the
\fxfopt\ ratios that nearby AGNs would have if they were observed at a
redshift of $z = 0.3$.  Their analysis revealed that many such objects would
have low \fxfopt\ ratios and modest X-ray luminosities, similar to the
optically bright, X-ray--faint sources (OBXFs; Hornschemeier et al.\ 2001,
2003) that have been detected in the CDF-N.  Spectroscopically, the OBXFs
appear to be quiescent, and Peterson et al.\ have reasoned that many could
harbor normal Seyfert~2 nuclei if host-galaxy dilution is significant in
their ground-based optical spectra.  Our results support this conclusion.
In Figure~5, it is clear that redshift effects on \fxfi\ are slight at $z =
0.3$.  All but a few of the UW89 objects would be detectable at $z \approx
0.3$ in the CDF-N, so the UW89 \fxfi\ distribution at that redshift would
look much like the $z = 0$ distribution shown in Figure~6, shifted by only
$\sim +0.3$ in log~\fxfi .  A significant number of the UW89 galaxies would
therefore have log~\fxfopt\ $< -2$, similar to the OBXFs in the CDF-N.  In
addition, the low-\fxfopt\ objects in the sample would have X-ray luminosities
in the range of normal galaxies ($\sim 10^{41}$ ergs~s$^{-1}$ or less), and
many would have quiescent optical spectra (Moran et al.\ 2002).  Thus, as
Peterson et al.\ have suggested, a number of the OBXF objects could be
unrecognized Seyfert~2s.

\section{Summary and Conclusions}

To investigate the nature of the ``normal'' X-ray--luminous galaxies in the
CDF-N, we have obtained {\sl UBVRI\/} photometry and broadband X-ray data for
a distance-limited Seyfert~2 galaxy sample that broadly represents the local
luminosity function for absorbed AGNs.  From these data we have measured the
integrated fluxes of the galaxies, since this is what is normally derived
from multiwavelength observations of the distant objects detected in the
deep X-ray surveys.

We have selected a sample of absorbed AGNs from a well-defined portion of the
CDF-N for comparison to the local objects.  Using the redshift distribution
of the CDF-N sources, we have simulated the \fxfopt\ ratios that the UW89
objects would have if they were observed at modest redshift as part of the
CDF-N.  By including (1) the effects of redshift on flux measurements in fixed
observed-frame bands, and (2) the way the luminosity function of a given
population is sampled in a flux-limited survey like the CDF-N, we have shown
that nearby Seyfert~2s with strong nuclear emission lines are able to account
for the X-ray and optical properties of distant absorbed AGNs, despite the
fact that the latter often lack optical evidence for nuclear activity in
ground-based data.  The integrated spectra of UW89 galaxies indicate that
observational factors --- host-galaxy dilution, signal-to-noise ratio, and
wavelength coverage --- are capable of hiding the nuclear emission lines of
bona fide Seyfert~2s (Moran et al.\ 2002).  We conclude, therefore, that the
same factors provide the simplest explanation for the ``normal'' appearance
of many absorbed AGNs in the {\it Chandra\/} deep surveys.  Note that our
arguments are statistical --- it is certainly possible that some distant
absorbed AGNs appear to be normal because they are located in edge-on host
galaxies, or because they have unusually high amounts of nuclear obscuration.
In general, though, we have been unable to identify differences between the
nearby and distant populations of absorbed AGNs that cannot be attributed
to host-galaxy dilution.  Until we do, it seems unnecessary to invoke the
existence of a significant new class of X-ray--bright, optically normal
galaxies (XBONGs; Comastri et al.\ 2002) that differ from nearby Seyfert~2s
in some fundamental way.

The problem with the X-ray--luminous ``normal'' galaxies may be mainly a
matter of perception.  In Figure~10, we have plotted the integrated spectra
of two galaxies from the UW89 sample, Mrk~3 and NGC~788 (Moran et al.\ 2002).
In most respects, these two AGNs are nearly identical: they have similar
X-ray luminosities and absorption column densities; optically, their
luminosities are comparable and both exhibit polarized broad emission lines;
and both reside in S0 host galaxies at a distance of $d \approx 54$ Mpc.
However, as Figure~10 illustrates, a wide range of line strengths exists
among ``real'' Seyfert~2s.  Mrk~3 would be easily recognized as an AGN at
moderate redshifts, whereas NGC~788 would not.  The main difficulty with
the deep X-ray survey results may lie with an expectation that the average
Seyfert 2 resembles Mrk~3, when in fact NGC~788 is the more typical object.

\acknowledgments
We would like to thank John Salzer for helpful discussions regarding Malmquist
effects in flux-limited surveys, Seth Cohen for help with the axis-ratio
measurements, Eve Armstrong for obtaining the optical images of NGC~2110,
Kaitlin Kratter for extensive help with the observing at the MDM 1.3-m, and
Mary Hui for assistance with the WIYN 0.9-m observing.  This work was supported
in part by NASA through a grant for {\sl HST\/} proposal \#09869 from the
Space Telescope Science Institute, which is operated the Association of
Universities for Research in Astronomy, Inc., under NASA contract NAS5-26555.

\clearpage
\begin{center}
\begin{deluxetable}{lccccc}
\tablewidth{0pt}
\tablecaption{Photometry Results}
\tablehead{\colhead{Galaxy} &
           \colhead{$U$} &
           \colhead{$B$} &
           \colhead{$V$} &
           \colhead{$R$} &
           \colhead{$I$}}
\startdata
MCG $-$05-18-002 & 14.12 & 13.52 & 11.92 & 11.28 & 10.49 \\
MCG $+$01-27-020 & 14.82 & 14.83 & 14.08 & 13.65 & 13.05 \\
Mrk 3            & 13.88 & 13.60 & 12.39 & 11.64 & 10.96 \\
Mrk 1066         & 14.50 & 14.17 & 13.19 & 12.56 & 11.83 \\
NCG 262	         & 13.68 & 13.67 & 12.84 & 12.28 & 11.72 \\
NCG 591	         & 14.15 & 14.00 & 13.18 & 12.62 & 11.96 \\
NGC 788	         & 13.45 & 13.02 & 12.05 & 11.50 & 10.80 \\
NGC 1358         & 13.50 & 13.06 & 12.09 & 11.48 & 10.83 \\
NGC 1667         & 12.90 & 12.79 & 12.03 & 11.46 & 10.82 \\
NGC 1685         & 14.58 & 14.29 & 13.40 & 12.73 & 12.16 \\
NGC 2110         & 14.50 & 13.22 & 11.83 & 11.06 & 10.18 \\
NGC 2273         & 12.86 & 12.65 & 11.64 & 11.02 & 10.31 \\
NGC 3081         & 13.22 & 12.96 & 12.05 & 11.51 & 10.87 \\
NGC 3982         & 12.10 & 12.18 & 11.59 & 11.14 & 10.57 \\
NGC 4117         & 14.09 & 13.84 & 13.00 & 12.43 & 11.78 \\
NGC 4388         & 11.86 & 11.72 & 10.96 & 10.45 & ~9.76 \\
NGC 4941         & 12.03 & 12.06 & 11.15 & 10.60 & ~9.86 \\
NGC 5347         & 13.23 & 13.17 & 12.46 & 11.93 & 11.14 \\
NGC 5695         & 13.82 & 13.49 & 12.66 & 12.09 & 11.31 \\
NGC 5929         & 12.76 & 12.99 & 12.05 & 11.45 & 10.72 \\
NGC 7672         & 14.95 & 14.76 & 13.94 & 13.40 & 12.50
\enddata
\vskip -0.3truein
\tablecomments{Uncorrected for Galactic extinction.}
\end{deluxetable}
\end{center}

\begin{center}
\begin{deluxetable}{lcccccc}
\tablewidth{0pt}
\tablecaption{Integrated Magnitudes from the Literature}
\tablehead{\colhead{Galaxy} &
           \colhead{$U$} &
           \colhead{$B$} &
           \colhead{$V$} &
           \colhead{$R$} &
           \colhead{$I$} &
           \colhead{Lit.\ data/template}}
\startdata
IC 3639	         & 13.34 & 13.01 & 12.23 & 11.87 & 11.22 & $BR$/Sbc \\ 
MCG $-$05-23-016 & 14.49 & 14.07 & 13.29 & 12.44 & 11.83 & $BR$/S0 \\
NCG 424	         & 14.18 & 13.76 & 12.91 & 12.38 & 11.77 & $BR$/S0 \\
NGC 1068         & ~9.70 & ~9.61 & ~8.87 & ~8.31 & ~7.66 & $UBV$/Sab \\
NGC 1386         & 12.42 & 12.09 & 11.23 & 10.76 & 10.15 & $UBVR$/S0 \\	
NGC 2992         & 13.54 & 13.14 & 12.18 & 11.62 & 10.97 & $UBV$/S0 \\
NGC 3281         & 13.12 & 12.70 & 11.72 & 11.17 & 10.56 & $BV$/S0 \\	
NGC 4507         & 13.05 & 12.92 & 12.07 & 11.70 & 11.05 & $UBVR$/Sab \\
NGC 5135         & 13.01 & 12.88 & 12.11 & 11.55 & 10.90 & $UBV$/Sab \\
NGC 5283         & 14.53 & 14.20 & 13.42 & 12.86 & 12.21 & $B$/Sab \\
NGC 5506         & 13.21 & 12.79 & 11.92 & 11.38 & 10.77 & $BV$/S0 \\
NGC 5643         & 10.89 & 10.74 & 10.00 & ~9.48 & ~8.87 & $UBV$/Sbc \\
NGC 5728         & 12.70 & 12.37 & 11.59 & 11.03 & 10.38 & $B$/Sab \\
NGC 6890         & 13.14 & 13.01 & 12.25 & 11.57 & 10.92 & $UBVR$/Sab \\
NGC 7172         & 13.24 & 12.85 & 11.91 & 11.15 & 10.54 & $UBVR$/S0 \\
NGC 7314         & 11.57 & 11.62 & 11.01 & 10.61 & ~9.99 & $UBVR$/Sbc \\
NGC 7582         & 11.62 & 11.37 & 10.62 & 10.06 & ~9.41 & $UBV$/Sab 
\enddata
\vskip -0.3truein
\tablecomments{Uncorrected for Galactic extinction.}
\end{deluxetable}
\end{center}

\begin{center}
\begin{deluxetable}{lccccc}
\tablewidth{0pt}
\tabletypesize \scriptsize
\tablecaption{Magnitudes Corrected for Galactic Extinction}
\tablehead{\colhead{Galaxy} &
           \colhead{$U$} &
           \colhead{$B$} &
           \colhead{$V$} &
           \colhead{$R$} &
           \colhead{$I$}}
\startdata
IC 3639      & {\it 12.96} & 12.71 & {\it 12.00} & 11.69       & {\it 11.09}\\
MCG $-$05-18-002  & 13.05  & 12.67 & 11.27       & 10.75       & 10.10\\
MCG $-$05-23-016  & {\it 13.90} & 13.60 & {\it 12.93} & 12.15 & {\it 11.62}\\
MCG $+$01-27-020  & 14.66  & 14.70 & 13.98       & 13.57       & 12.99\\
Mrk 3        & 12.86       & 12.79 & 11.77       & 11.14       & 10.60\\
Mrk 1066     & 13.79       & 13.60 & 12.75       & 12.21       & 11.57 \\
NCG 262      & 13.32       & 13.38 & 12.62       & 12.10       & 11.59 \\
NCG 424      & {\it 14.10} & 13.69 & {\it 12.86} & 12.34       & {\it 11.74}\\
NCG 591      & 13.90       & 13.80 & 13.03       & 12.50       & 11.87\\
NGC 788      & 13.30       & 12.91 & 11.96       & 11.43       & 10.75\\
NGC 1068     & ~9.52       & ~9.47 & ~8.76       & {\it ~8.22} & {\it ~7.60}\\
NGC 1358     & 13.16       & 12.79 & 11.88       & 11.31       & 10.71\\
NGC 1386     & 12.35       & 12.04 & 11.19       & 10.73       & {\it 10.13}\\
NGC 1667     & 12.46       & 12.45 & 11.76       & 11.24       & 10.67\\
NGC 1685     & 14.27       & 14.05 & 13.21       & 12.58       & 12.05\\
NGC 2110     & 12.46       & 11.60 & 10.56       & 10.06       & 9.45\\
NGC 2273     & 12.41       & 12.30 & 11.37       & 10.81       & 10.15\\
NGC 2992     & 13.21       & 12.88 & 11.98       & {\it 11.46} & {\it 10.85}\\
NGC 3081     & 12.92       & 12.72 & 11.87       & 11.37       & 10.77\\
NGC 3281     & {\it 12.60} & 12.29 & 11.40       & {\it 10.91} & {\it 10.37}\\
NGC 3982     & 12.02       & 12.12 & 11.55       & 11.11       & 10.54\\
NGC 4117     & 14.02       & 13.79 & 12.95       & 12.39       & 11.75\\
NGC 4388     & 11.68       & 11.58 & 10.85       & 10.36       & ~9.69\\
NGC 4507     & 12.52       & 12.50 & 11.75       & 11.44       & {\it 10.86}\\
NGC 4941     & 11.83       & 11.90 & 11.03       & 10.51       & ~9.79\\
NGC 5135     & 12.69       & 12.62 & 11.91       & {\it 11.39} & {\it 10.78}\\
NGC 5283     & {\it 13.98} & 13.76 & {\it 13.08} & {\it 12.59} & {\it 12.01}\\
NGC 5347     & 13.11       & 13.08 & 12.45       & 11.87       & 11.09\\
NGC 5506     & {\it 12.89} & 12.53 & 11.72       & {\it 11.22} & {\it 10.65}\\
NGC 5643     & ~9.97       & 10.01 & ~9.44       & {\it ~9.03} & {\it ~8.54}\\
NGC 5695     & 13.73       & 13.41 & 12.60       & 12.05       & 11.28\\
NGC 5728     & {\it 12.59} & 12.28 & {\it 11.52} & {\it 10.98} & {\it 10.34}\\
NGC 5929     & 12.63       & 12.89 & 11.97       & 11.38       & 10.67\\
NGC 6890     & 12.92       & 12.84 & 12.12       & 11.46       & {\it 10.84}\\
NGC 7172     & 13.10       & 12.74 & 11.82       & 11.08       & {\it 10.49}\\
NGC 7314     & 11.45       & 11.53 & 10.94       & 10.55       & {\it ~9.95}\\
NGC 7582     & 11.54       & 11.31 & 10.57       & {\it 10.02} & {\it ~9.38}\\
NGC 7672     & 14.56       & 14.45 & 13.70       & 13.21       & 12.36
\enddata
\end{deluxetable}
\end{center}

\begin{center}
\begin{deluxetable}{lcccccccccc}
\tablewidth{0pt}
\rotate
\tabletypesize \footnotesize
\tablecaption{X-ray Spectral Parameters and Fluxes}
\tablehead{\colhead{Galaxy} &
           \colhead{$d$(Mpc)} &
           \colhead{$(N_{\rm H})_1$} &
           \colhead{$\Gamma_1$} &
           \colhead{$(N_{\rm H})_2$} &
           \colhead{$\Gamma_2$} &
           \colhead{$E_{K\alpha}$} &
           \colhead{$\sigma_{K\alpha}$} &
           \colhead{EW$_{K\alpha}$} &
           \colhead{$F$(0.5--2 keV)} &
           \colhead{$F$(2--8 keV)}}
\startdata
IC 3639          & 43.7 &    0    & 1.78 & 2.42E23 & 1.70 &  ... &  ...  &  ... &  1.30E-13 &  2.43E-13\\
MCG $-$05-18-002 & 23.1 &    0    & 1.78 & 2.42E23 & 1.70 &  ... &  ...  &  ... &  2.07E-13 &  5.28E-13\\
MCG $-$05-23-016 & 33.3 &   ...   &  ... & 1.50E22 & 1.80 & 6.35 & 0.39  & 0.29 &  8.25E-12 &  7.21E-11\\
MCG $+$01-27-020 & 46.8 &    0    & 1.78 & 2.42E23 & 1.70 &  ... &  ...  &  ... &  3.40E-14 &  4.21E-13\\
Mrk 3            & 54.0 &    0    & 2.05 & 4.74E23 & 1.83 & 6.31 & 0.10  & 0.73 &  8.26E-13 &  2.77E-12\\
Mrk 1066         & 48.1 &    0    & 1.78 & 2.42E23 & 1.70 &  ... &  ...  &  ... &  1.62E-13 &  3.92E-13\\
NCG 262          & 60.1 &    0    & 1.77 & 1.77E23 & 1.77 & 6.16 & 0.10  & 0.14 &  1.38E-13 &  3.56E-12\\
NCG 424          & 46.6 &    0    & 1.61 & 2.42E23 & 1.61 & 6.41 & 0.10  & 1.49 &  2.14E-13 &  1.04E-12\\
NCG 591          & 60.7 &    0    & 1.78 & 2.42E23 & 1.70 &  ... &  ...  &  ... &  3.76E-14 &  2.90E-13\\
NGC 788          & 54.4 &    0    & 1.15 & 4.80E23 & 1.67 & 6.23 & 0.24  & 0.39 &  8.13E-14 &  3.48E-12\\
NGC 1068         & 14.4 & 2.90E21 & 5.59 & 3.16E21 & 1.28 & 6.52 & 0.30  & 3.26 &  1.02E-11 &  5.19E-12\\
NGC 1358         & 53.8 &    0    & 1.78 & 2.42E23 & 1.70 &  ... &  ...  &  ... &  5.14E-14 &  4.27E-13\\
NGC 1386         & 16.9 & 2.04E21 & 3.47 & 1.94E23 & 1.96 & 6.40 & 0.10  &    0 &  4.90E-13 &  4.46E-13\\
NGC 1667         & 60.7 &    0    & 1.78 & 2.42E23 & 1.70 &  ... &  ...  &  ... &  2.81E-14 &  1.27E-13\\
NGC 1685         & 60.4 &    0    & 1.78 & 2.42E23 & 1.70 &  ... &  ...  &  ... &  3.38E-14 &  8.31E-14\\
NGC 2110         & 29.1 & 5.80E20 & 1.61 & 3.61E22 & 1.61 & 6.26 & 0.51  & 0.41 &  1.20E-12 &  2.39E-11\\
NGC 2273         & 28.4 &    0    & 1.95 & 4.41E23 & 1.95 & 6.36 & 0.11  & 1.38 &  2.25E-13 &  9.69E-13\\
NGC 2992         & 30.5 &    0    & 4.58 & 9.74E21 & 1.61 & 6.46 & 0.47  & 1.59 &  1.30E-12 &  3.99E-12\\
NGC 3081         & 32.5 &    0    & 2.01 & 5.36E23 & 1.58 & 6.22 & 0.13  & 0.41 &  2.12E-13 &  3.75E-12\\
NGC 3281         & 42.7 &    0    & 1.85 & 6.02E23 & 1.64 & 6.30 & 0.10  & 0.99 &  2.72E-13 &  2.25E-12\\
NGC 3982         & 27.2 &    0    & 1.78 & 2.42E23 & 1.70 &  ... &  ...  &  ... &  1.03E-13 &  3.22E-13\\
NGC 4117         & 17.0 & 2.21E21 & 1.60 & 3.99E23 & 1.60 & 6.43 & 0.10  & 0.11 &  7.62E-14 &  1.31E-12\\
NGC 4388         & 16.8 &    0    & 1.50 & 3.98E23 & 1.96 & 6.34 & 0.16  & 0.62 &  5.41E-13 &  4.86E-12\\
NGC 4507         & 47.2 &    0    & 1.78 & 3.02E23 & 1.53 & 6.23 & 0.23  & 0.30 &  5.41E-13 &  1.73E-11\\
NGC 4941         & ~6.4 & 1.18E21 & 1.61 & 9.13E23 & 1.61 & 6.35 & 0.10  & 0.64 &  1.08E-13 &  8.26E-13\\
NGC 5135         & 54.9 &    0    & 2.88 & 3.84E23 & 1.80 & 6.34 & 0.10  & 1.44 &  5.36E-13 &  4.65E-13\\
NGC 5283         & 41.4 & 1.10E21 & 3.28 & 1.06E23 & 1.60 & 6.40 & 0.10  &    0 &  3.91E-14 &  1.14E-12\\
NGC 5347         & 36.7 &    0    & 1.78 & 2.42E23 & 1.70 &  ... &  ...  &  ... &  5.16E-14 &  1.94E-13\\
NGC 5506         & 28.7 & 1.11E21 & 2.20 & 3.02E22 & 1.82 & 6.39 & 0.23  & 0.22 &  3.81E-12 &  6.03E-11\\
NGC 5643         & 16.9 &    0    & 1.90 & 1.76E23 & 1.64 & 6.34 & 0.10  & 2.73 &  4.49E-13 &  1.12E-12\\
NGC 5695         & 56.4 &    0    & 1.78 & 2.42E23 & 1.70 &  ... &  ...  &  ... &  5.05E-14 &  1.23E-13\\
NGC 5728         & 42.2 & 3.63E20 & 2.07 & 7.17E23 & 1.67 & 6.33 & 0.10  & 1.12 &  9.22E-14 &  1.00E-12\\
NGC 5929         & 38.5 & 5.16E21 & 1.70 & 2.77E23 & 1.70 & 6.19 & 0.10  & 0.35 &  8.06E-14 &  1.40E-12\\
NGC 6890         & 31.8 &    0    & 1.78 & 2.42E23 & 1.70 &  ... &  ...  &  ... &  3.99E-14 &  1.69E-13\\ 
NGC 7172         & 33.9 &    0    & 1.69 & 9.62E22 & 1.82 & 6.44 & 0.10  & 0.20 &  1.93E-13 &  1.01E-11\\
NGC 7314         & 18.3 &   ...   &  ... & 7.62E21 & 1.95 & 6.24 & 0.77  & 0.45 &  8.36E-12 &  3.42E-11\\
NGC 7582         & 17.6 &    0    & 1.68 & 1.23E23 & 1.82 & 6.24 & 0.10  & 0.18 &  4.51E-13 &  1.23E-11\\
NGC 7672         & 53.5 &    0    & 1.78 & 2.42E23 & 1.70 &  ... &  ...  &  ... &  4.80E-14 &  9.08E-14
\enddata
\vskip -0.1truein
\tablecomments{Column densities ($N_{\rm H}$) are in units of atoms cm$^{-2}$.
Iron line centroids ($E_{K\alpha}$), energy widths ($\sigma_{K\alpha}$),
and equivalent widths (EW$_{K\alpha}$) are all in units of keV.  Fluxes are
in units of erg~cm$^{-2}$~s$^{-1}$.}
\end{deluxetable}
\end{center}

\begin{center}
\begin{deluxetable}{lc}
\tablewidth{0pt}
\tabletypesize \footnotesize
\tablecaption{Host Galaxy Axis Ratios}
\tablehead{\colhead{Galaxy} &
           \colhead{$b/a$}}
\startdata
MCG $-$05-23-016~~~~~~~~~~~~~~~~~~~~~ & 0.45\\
Mrk 3            & 0.97\\
NCG 262          & 0.79\\
NCG 424          & 0.33\\
NCG 788          & 0.76\\
NCG 2110         & 0.80\\
NCG 2992         & 0.39\\
NCG 3081         & 0.57\\
NCG 3281         & 0.43\\
NCG 4388         & 0.28\\
NCG 4507         & 0.86\\
NCG 5283         & 0.91\\
NCG 5506         & 0.32\\
NCG 5728         & 0.74\\
NCG 5929         & 0.78\\
NCG 7172         & 0.52\\
NCG 7314         & 0.39\\
NCG 7582         & 0.46
\enddata
\vskip -0.1truein
\tablecomments{Galaxies included here are those that would be detected as
absorbed AGNs in the CDF-S at $z \ge 0.5$.  NGC~5929 is interacting with
NGC~5920; the axis ratio listed is for NGC~5929 alone.}
\end{deluxetable}
\end{center}

\begin{figure}
\begin{center}
\epsscale{1.0}
\centerline{\plotone{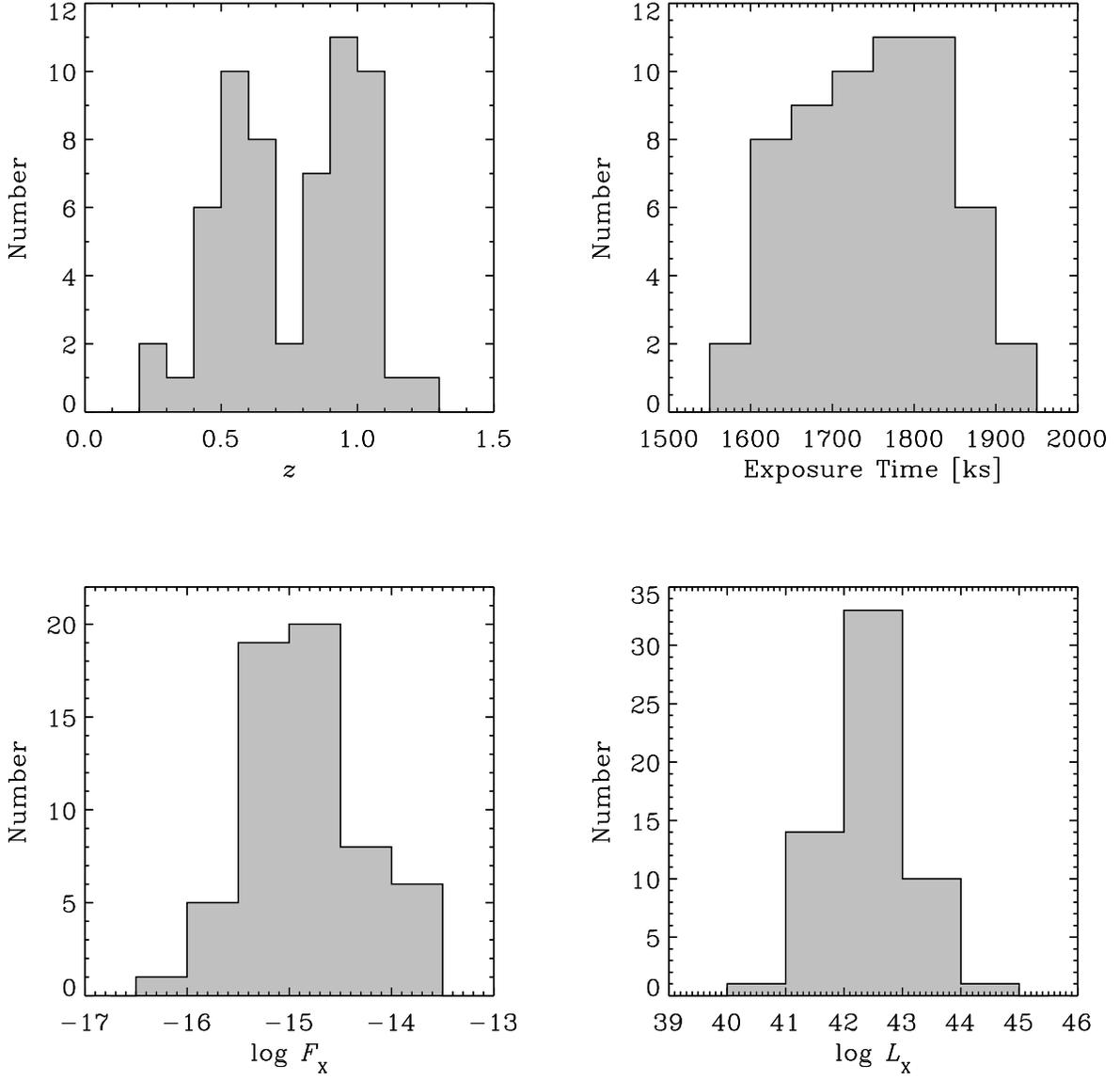}}
\vskip -0.1truein
\caption{Distributions of the redshifts, ACIS-I exposure times, 2--8 keV
fluxes, and observed-frame 2--8 keV luminosities of the absorbed AGNs in
the CDF-N that meet the selection criteria described in \S~2.}
\end{center}
\end{figure}

\begin{figure}
\begin{center}
\epsscale{1.0}
\centerline{\plottwo{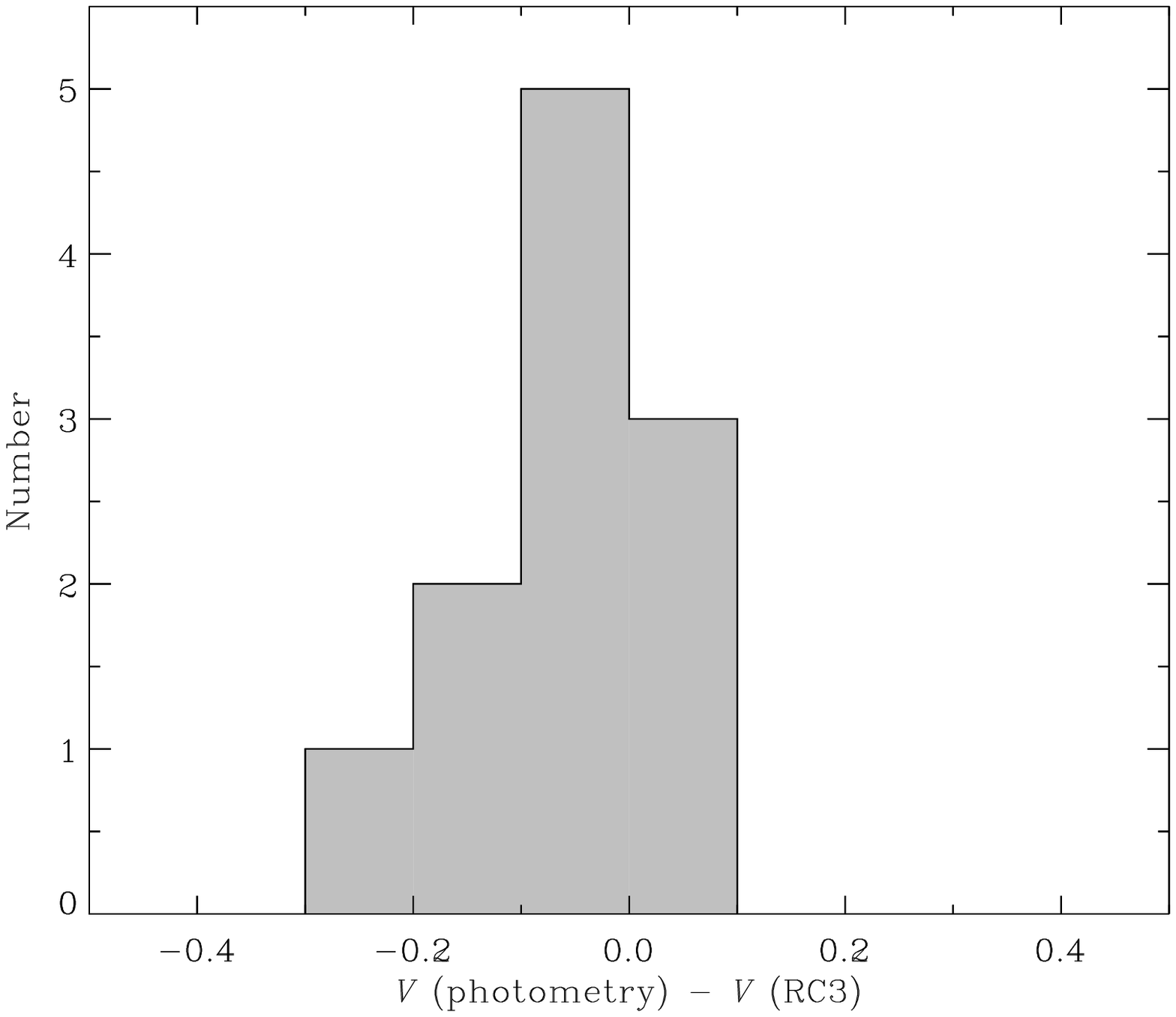}{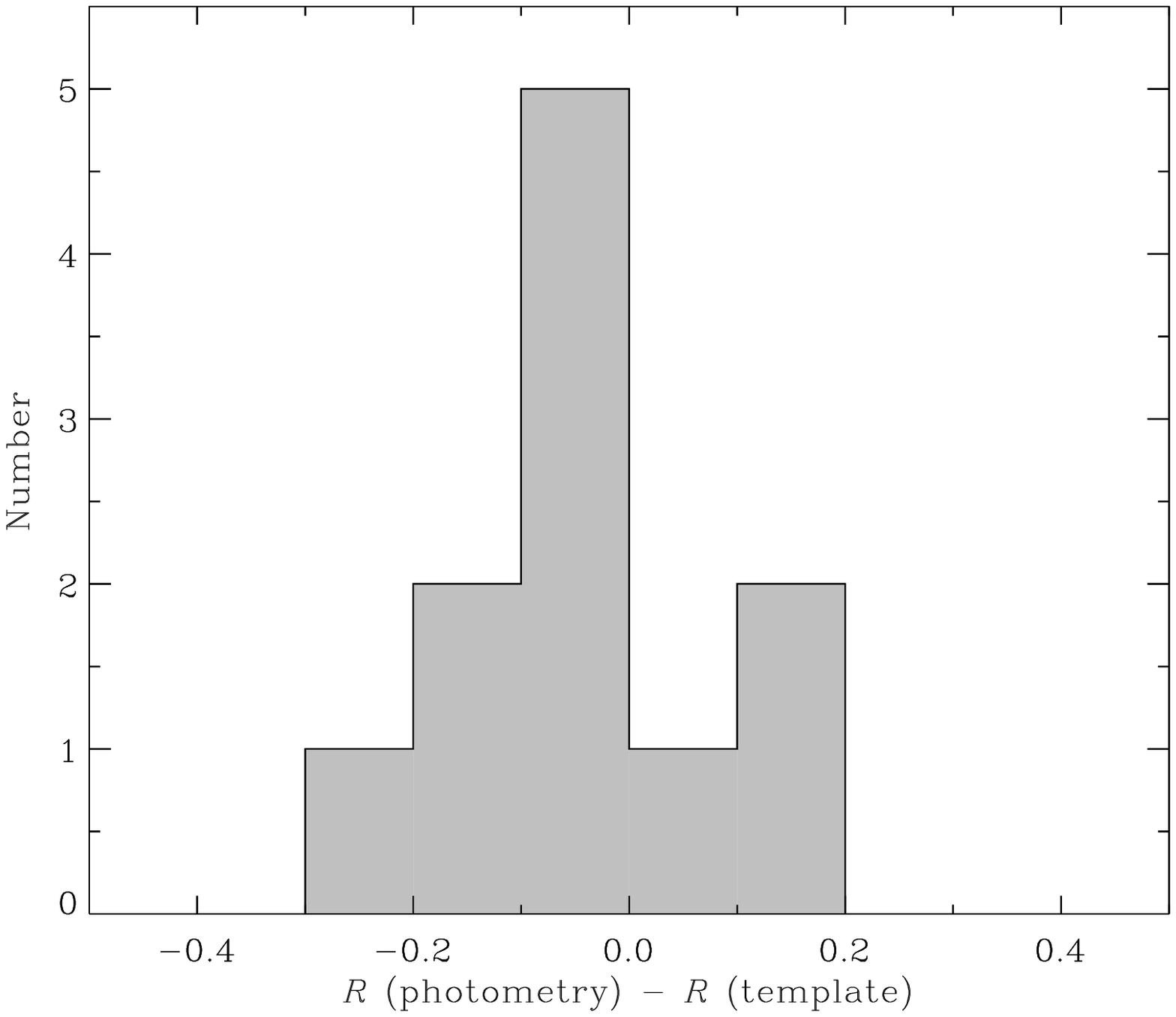}}
\vskip -0.7truein
\caption{({\it left panel}) Comparison of the integrated $V$ magnitudes of
11 objects from our CCD photometry with those from photoelectric measurements
published in the RC3.  ({\it right panel}) Integrated $R$ magnitudes of the
same objects from our photometry, compared with those extrapolated from the 
RC3 $V$ magnitude using a Fukugita et al.\ (1995) galaxy color template.
In both cases, the mean and dispersion of the magnitude differences are
consistent with the uncertainties in the RC3 measurements.}
\end{center}
\end{figure}

\begin{figure}
\begin{center}
\epsscale{1.0}
\centerline{\plotone{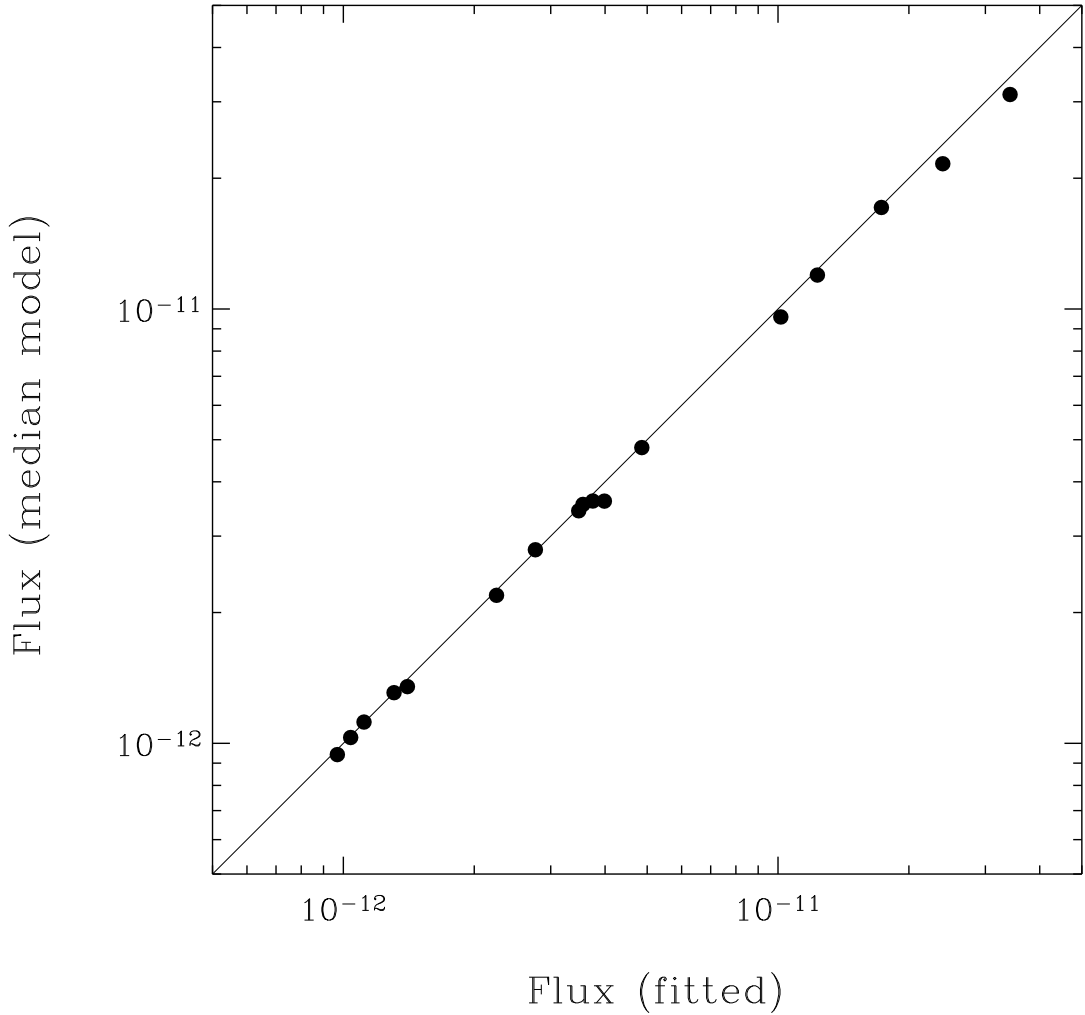}}
\vskip -0.5truein
\caption{Comparison of the 2--8 keV fluxes of the well-detected UW89 objects
obtained directly from spectral modeling with those obtained by scaling the
median model to agree with the measured (4--10 keV)/(1--4 keV) counts ratio.
For most sources, the two fluxes differ by only a few percent, suggesting
that the application of the median model provides accurate flux estimates
for the 13 weakly detected UW89 objects.}
\end{center}
\end{figure}

\begin{figure}
\begin{center}
\epsscale{0.9}
\centerline{\plotone{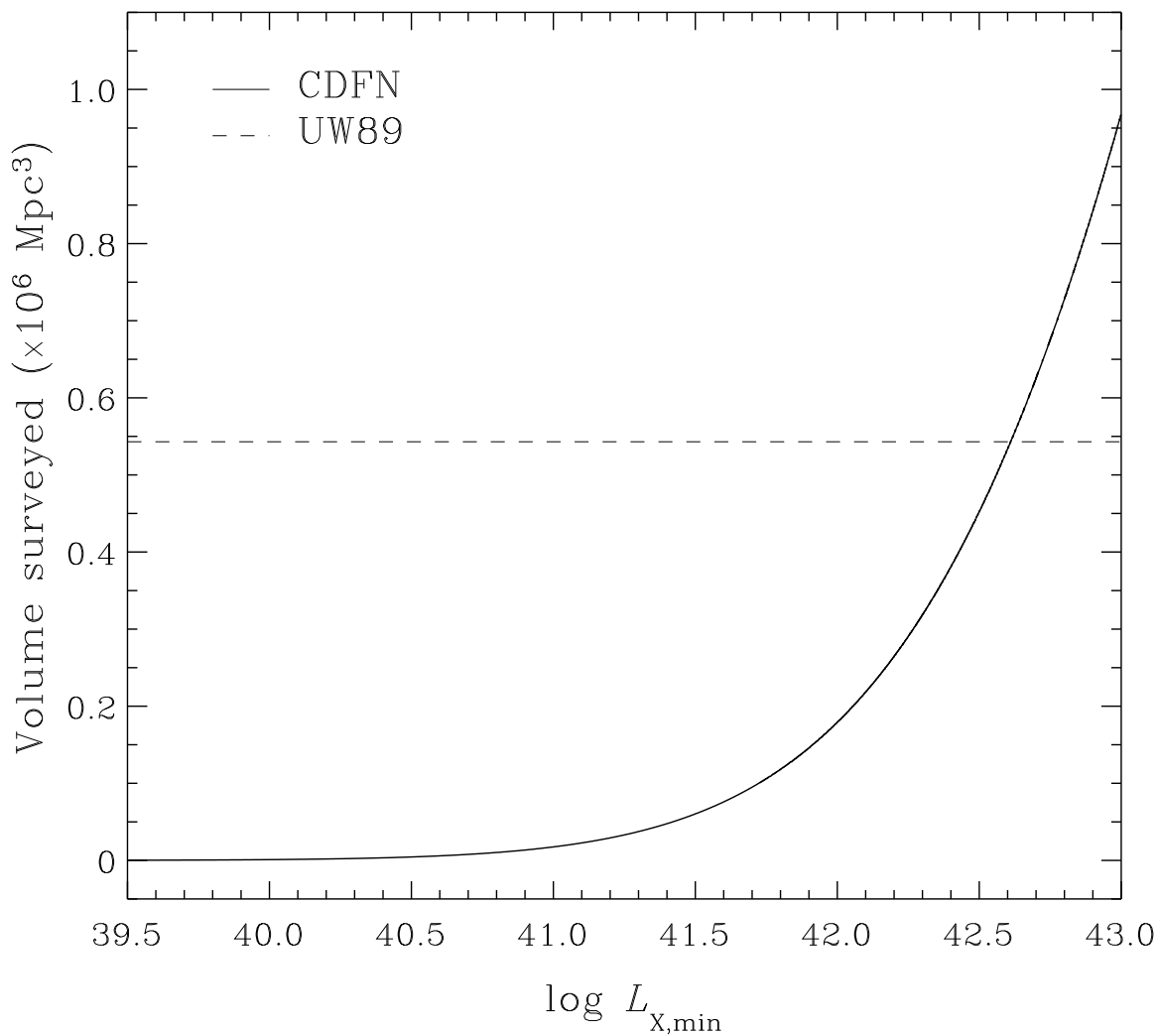}}
%\vskip -1.5truein
\caption{The volume searched in the CDF-N vs.\ minimum detectable luminosity in
the 2--8 keV band.  The volume is calculated assuming the flux limit and solid
angle that correspond to an ACIS-S exposure time of 1500~ks.  Also plotted is
the (fixed) volume covered by the UW89 sample of nearby Seyfert~2 galaxies.}
\end{center}
\end{figure}

\begin{figure}
\begin{center}
\epsscale{0.85}
\centerline{\plotone{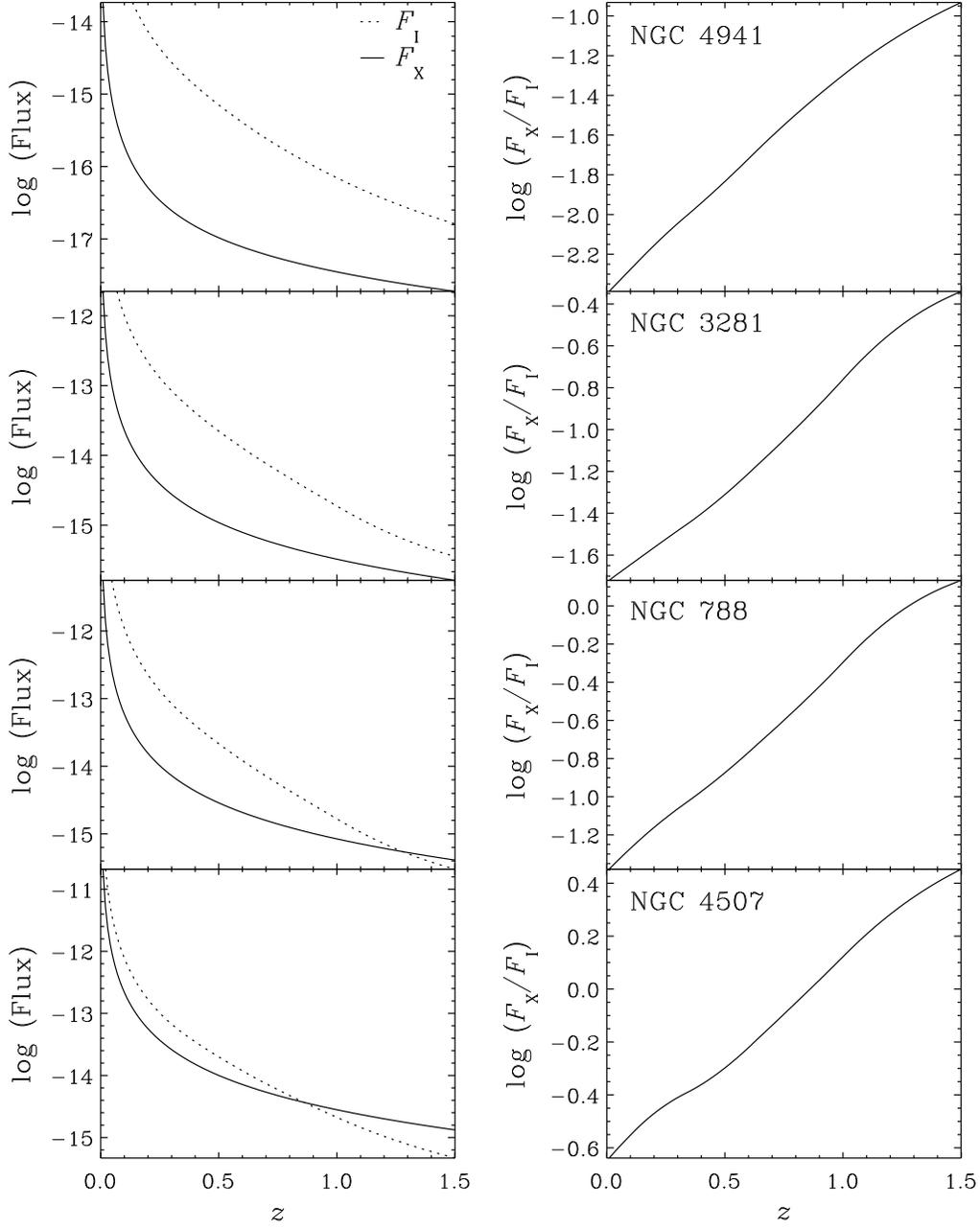}}
\vskip -1.0truein
\caption{({\it left}) Fluxes in the observed 2--8 keV and $I$ bands vs.\
redshift, and ({\it right}) \fxfi\ vs.\ redshift for four UW89 Seyfert~2s
that span a wide range of \fxfi\ at $z = 0$.  The \fxfi\ ratios of these
objects increase dramatically by factors of 15--35 as the redshift increases
from $z = 0$ to $z = 1.5$.}
\end{center}
\end{figure}

\begin{figure}
\begin{center}
\epsscale{0.9}
\centerline{\plotone{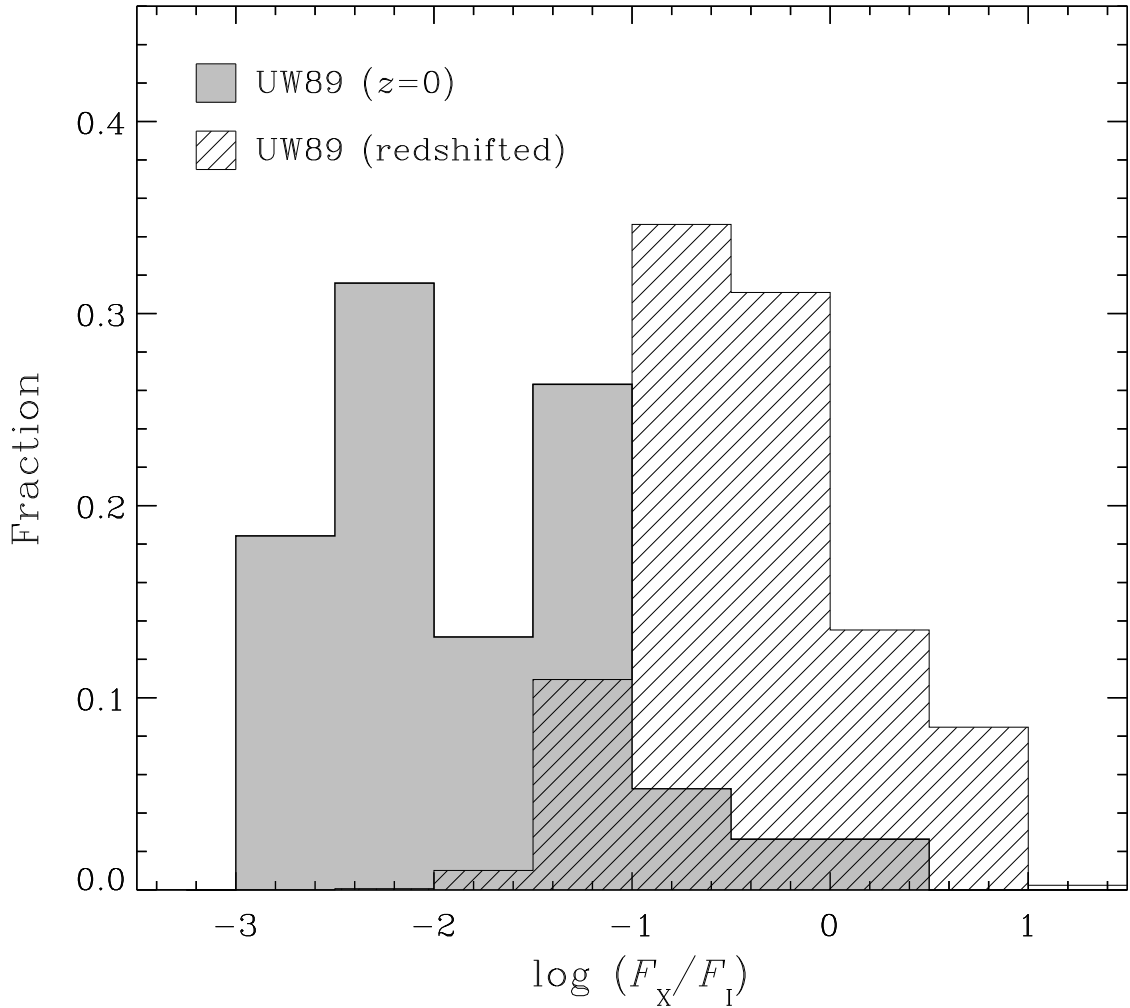}}
\vskip -0.0truein
\caption{Intrinsic ($z = 0$) \fxfi\ distribution for the UW89 sample, and
the distribution that would be obtained if the same sources were observed
in the CDF-N.  The dramatic transformation of the flux-ratio distribution
results from a combination of the redshift effects displayed in Fig.~5 and
the Malmquist effects (see \S~4.2) that arise because of the flux-limited
nature of the CDF-N.}
\end{center}
\end{figure}

\begin{figure}
\begin{center}
\epsscale{1.0}
\centerline{\plotone{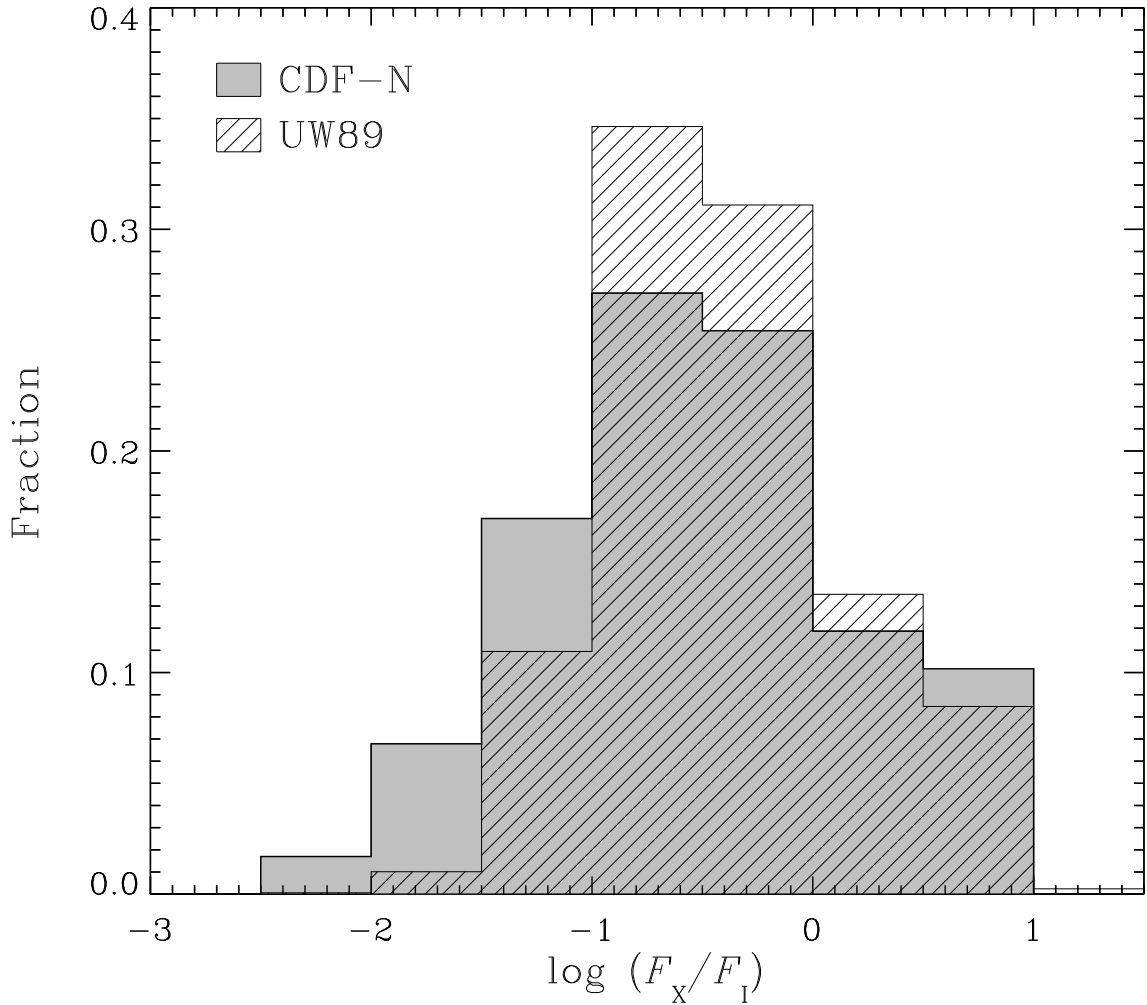}}
\vskip -1.7truein
\caption{Distribution of the observed-frame 2--8~keV/$I$-band flux ratio
for absorbed AGNs in the CDF-N, compared to the \fxfi\ ratios that the UW89
Seyfert~2 galaxies would have if they were observed in the CDF-N.  The
similarity of the distribuitions suggests that nearby Seyfert~2s and distant
absorbed AGNs do not differ in some fundamental way, despite the fact that
most of the latter lack spectroscopic evidence of activity in ground-based
optical observations.}
\end{center}
\end{figure}

\begin{figure}
\begin{center}
\epsscale{1.0}
\centerline{\plotone{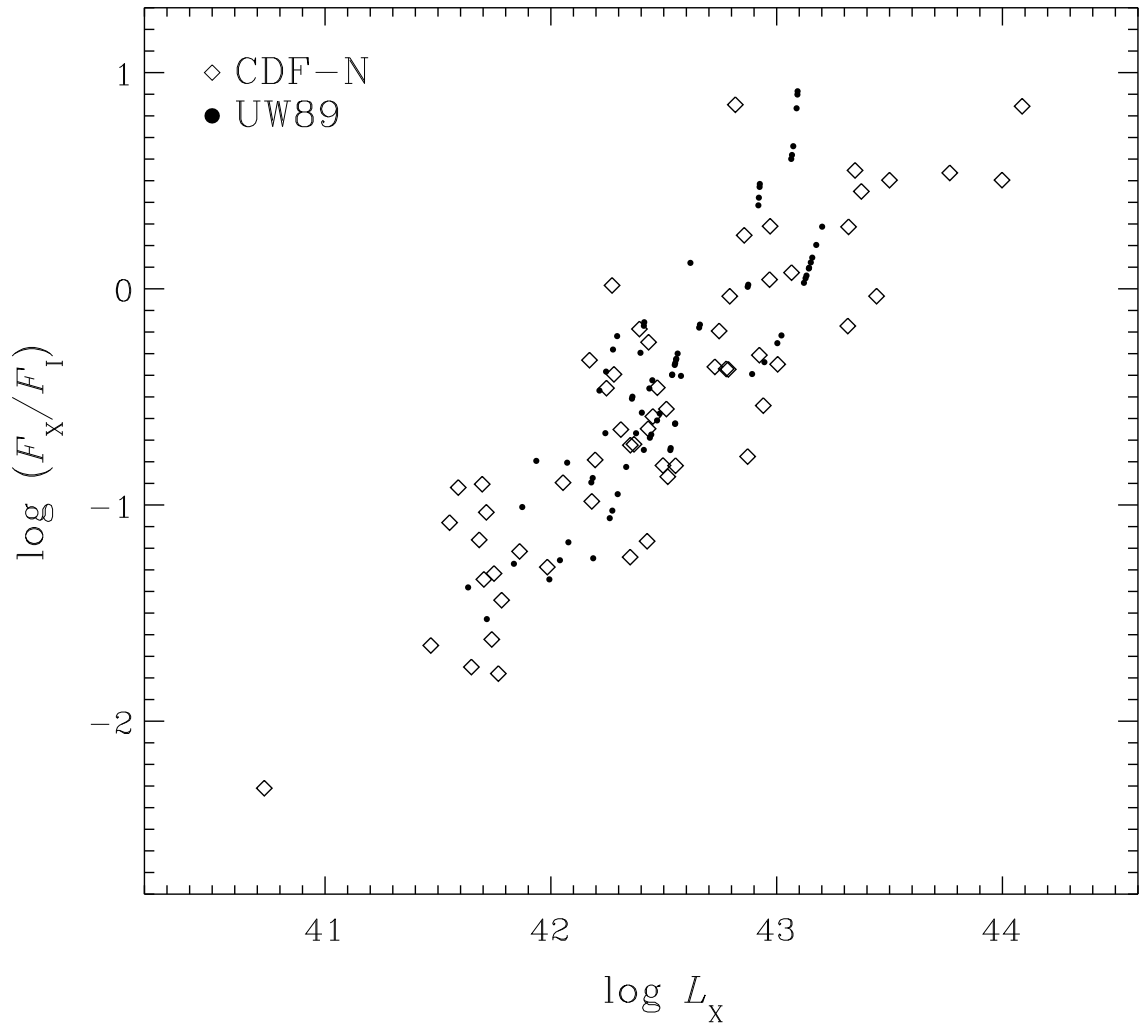}}
\vskip -1.7truein
\caption{\fxfi\ ratio for the CDF-N absorbed AGN sample, as a function of
observed 2--8 keV luminosity. Also plotted are points associated with UW89
galaxies from a simulation (see \S~4.1) consisting of 75 successful trials.
Overall, the simulated UW89 Seyfert~2 sample provides a close match to the
CDF-N objects over a broad range of luminosities and \fxfopt\ ratios.}
\end{center}
\end{figure}

\begin{figure}
\begin{center}
\epsscale{0.9}
\centerline{\plotone{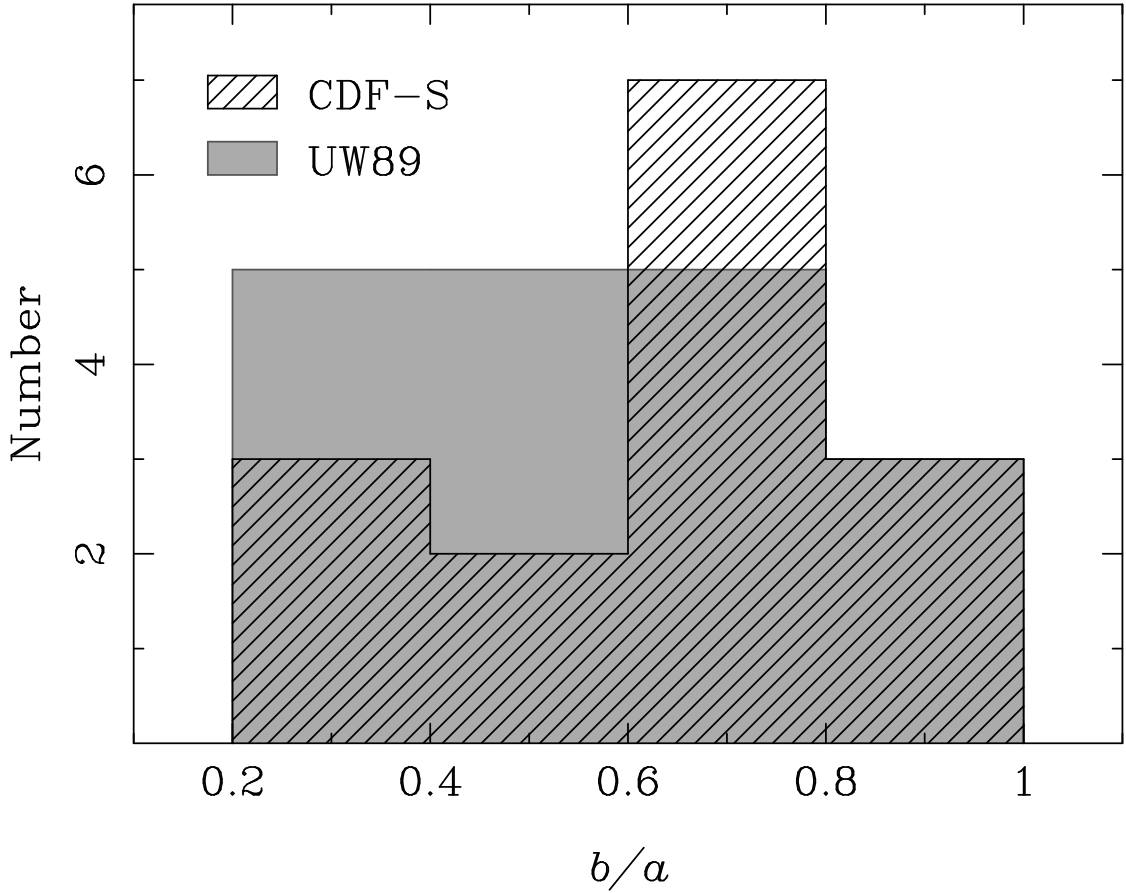}}
\vskip -0.0truein
\caption{Distribution of host galaxy axis ratios for subsets of the CDF-S and
UW89 samples.  The CDF-S sources are limited to those in the $0.5 \le z \le
0.8$ subsample of Rigby et al.\ (2006) that meet our criteria for absorbed
AGNs.  The UW89 objects included here are those that would be detected as
absorbed AGNs at $z \ge 0.5$ in the CDF-S.
The similarity of the CDF-S and UW89 axis-ratio distributions, and the fact
that the UW89 galaxies have strong nuclear emission lines, suggest that
inclination effects are not the main reason the CDF-S objects appear
optically inactive in ground-based spectra.}
\end{center}
\end{figure}

\begin{figure}
\begin{center}
\epsscale{1.0}
\centerline{\plotone{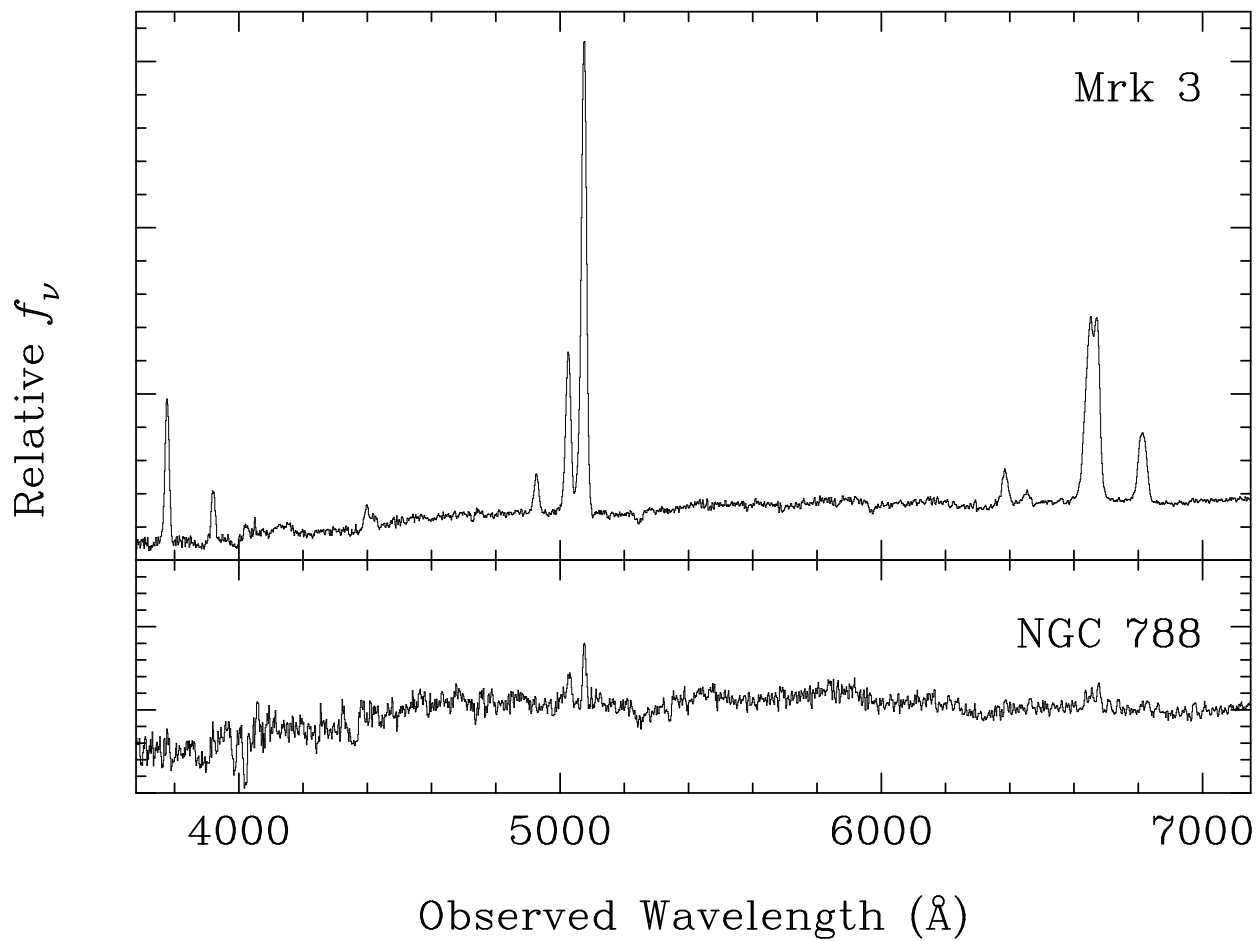}}
\vskip 0.0truein
\caption{Integrated spectra of two UW89 Seyfert~2 galaxies from the study
of Moran et al.\ (2002), plotted at the same scale.  Although the objects
are nearly identical in most respects, the strengths of their nuclear
emission lines are very different.  Objects like NGC~788, which is more
typical of nearby Seyfert~2s than Mrk~3 is, would not be recognized as
AGNs at $z > 0.5$ in the {\it Chandra\/} deep surveys.}
\end{center}
\end{figure}

\end{document}